\documentclass[a4paper,fleqn]{cas-sc}

\usepackage[authoryear,longnamesfirst]{natbib}

\def\tsc#1{\csdef{#1}{\textsc{\lowercase{#1}}\xspace}}
\tsc{WGM}
\tsc{QE}

\usepackage[english]{babel}
\usepackage{natbib}

\usepackage{amsmath}
\usepackage{amssymb}
\usepackage{graphicx}

\usepackage[left]{lineno}

\usepackage{overpic}
\usepackage{dynamo_symbols}
\newcommand{\dd}{\mathrm{d}}

\newcommand{\rs}{r_\mathrm{s}}
\newcommand{\ri}{r_\mathrm{i}}

\newcommand{\ro}{r_\mathrm{o}}
\newcommand{\hf}{H_\mathrm{F}}
\newcommand{\hs}{H_\mathrm{s}}
\newcommand{\dcsdr}{C_{\mathrm{s}}^{\prime}}
\newcommand{\dTsdr}{T_{\mathrm{i}}^{\prime}}
\newcommand{\dcidr}{C_{\mathrm{i}}^{\prime}}

\usepackage[]{siunitx}
\DeclareSIUnit\year{yr}
\DeclareSIUnit\wt{wt}
\DeclareSIUnit\gforce{\text{$g$}}

\begin{document}
\let\WriteBookmarks\relax
\def\floatpagepagefraction{1}
\def\textpagefraction{.001}

\title [mode = title]{Geomagnetic signatures of the slurry F-layer inferred from dynamo simulations}



\shorttitle{Geomagnetic signatures of the slurry F-layer inferred from dynamo simulations}   

\shortauthors{}  




%

\author[1]{Ludovic Huguet}[orcid=0000-0002-5532-6302]
\credit{Formal analysis; Investigation; Validation; Visualization; Writing – review and editing}
\cormark[1]

\ead{l.g.huguet@leeds.ac.uk}


\author[1]{Thomas Frasson}
\credit{Formal analysis; Investigation; Validation;  Visualization; Writing – review and editing}
\author[1]{Souvik Naskar}
\credit{Formal analysis; Investigation; Validation;  Visualization; Writing – review and editing}
\author[1]{Stephen J. Mason}
\credit{Formal analysis; Investigation; Validation;  Writing – review and editing}
\author[1]{Andrew T. Clarke}
\credit{Formal analysis; Investigation; Validation;  Writing – review and editing}
\author[1]{Hannah F. Rogers}
\credit{Formal analysis; Investigation; Validation;   Writing – review and editing}
\author[1]{Jonathan E. Mound}
\credit{Validation; Writing – review and editing}
\author[1]{Christopher J. Davies}
\credit{Conceptualization; Formal analysis; Investigation; Validation; Writing – original draft, review and editing }

\affiliation[1]{organization={School of Earth and Environment, University of Leeds},
            city={Leeds},
            postcode={LS2 9JT}, 
            country={UK}}


\begin{abstract}
Seismic observations indicate that the lowermost portion of Earth's liquid core is density stratified. The existence of this so-called F-layer challenges classical theories of core dynamics, where the geodynamo process that generates Earth's main magnetic field is assumed to be powered by heat and light element release at the inner core boundary. The seismically-inferred thickness, density, and velocity anomaly can be reproduced by a dynamical model that represents the F-layer as a two-phase two-component slurry on the liquidus, with a ``snow'' of solid iron particles falling through a quasi-static iron-oxygen liquid. Here, we present the first fluid dynamical simulations of thermochemically driven rotating convection and dynamo action that include a simple representation of the stratified slurry F-layer at the base of the spherical shell geometry. We show that the F-layer can create a barrier to columnar quasi-geostrophic flow, which is expressed near the core surface as a migration of peak radial and azimuthal flow speeds to lower latitudes as the thickness and stratification strength increase. In dynamo simulations, this effect induces polar minima in the radial magnetic field at the outer boundary ($B_r$) that strengthen and deepen with increasing stratification, and peaks in latitudinal profiles of $B_r$ moving to lower latitudes with reduced temporal variability. The geomagnetic signature of the F-layer is most prominent in time-averaged $B_r$, when resolved to at least spherical harmonic degree 5, and a trend of increasingly negative zonal degree 3 and 5 Gauss coefficients as the F-layer thickness and stratification strength increase. Our results suggest that an F-layer thickness of 600~km is incompatible with geomagnetic observations and favour weak stratification (normalised Brunt-Väisälä frequency $<1$) and a layer $<400$~km thick. 
\end{abstract}


\begin{highlights}
\item We examine the effect of a stably stratified F-layer at the bottom of the Earth's core in thermochemical convection and dynamo simulations.
\item The stably stratified F-layer creates a barrier to the flow near the tangent cylinder, which results in a migration of the flow structures associated with downwellings near the tangent cylinder toward the equator.
\item Our dynamo simulations suggest that the signature of the F-layer is visible in the radial magnetic field at the core-mantle boundary. 
\item Combining our results and geomagnetic observations gives new constraints on the structure of the F-layer: weakly stratified and thinner than 400~km.
\end{highlights}


\maketitle

\section{Introduction}

The F-layer is a region at the base of Earth's liquid outer core in which the gradient of compressional wave velocity departs from that of the Preliminary Reference Earth Model (PREM) \citep{dziewonski1981preliminary, souriau1991velocity,kennett1995constraints,ohtaki2018seismological, adam2018observation, ohtaki2025seismological}. The thickness of the layer is uncertain, but is usually taken to be between $100$ and $400$~km \citep{gubbins2008thermochemical}. The anomalous velocity gradient in the F-layer is usually assumed to reflect an increase in density with depth compared to PREM \citep{gubbins2008thermochemical,wasek2023}, and since PREM follows approximately an adiabatic profile, this implies that the F-layer is stably stratified. The stratification strength is also poorly constrained, but modelling suggests a normalised Brunt-Väisälä frequency above 1, indicating strong stratification \citep{gubbins2008thermochemical, wong2021regime}. The existence of a stratified F-layer presents a challenge to the classic model of core dynamics, where the power to drive the dynamo process that maintains the geomagnetic field derives primarily from the release of heat and light material at the inner core boundary (ICB). Reconciling this model of core dynamics with the existence of an F-layer thus requires an explanation of how heat and light elements can pass through the stable region without disturbing the stratification. 

Several mechanisms have been proposed to explain the existence of a stratified F-layer \citep[see reviews in ][]{wong2021regime, wilczynski2025two}. Here, we focus on the physical picture described by \citet{wilczynski2025two}, which refines earlier work by \citet{gubbins2008thermochemical}, \citet{wong2018boussinesq}, and \citet{wong2021regime}, where the F-layer is represented as a two-phase two-component slurry on the liquidus. The basic physical picture is that the stratification arises from a radial gradient in light element concentration, with the layer relatively depleted in light elements at its base compared to its top. This compositional stability owes its origin to barodiffusion and outward advection of light material in the liquid, driven by mass conservation between falling solid and rising liquid. Compositional stability overcomes the thermally destabilising conditions driven by heat release at the ICB. While this physical picture is still incomplete, particularly concerning the uncertain processes of solid nucleation \citep[e.g.][]{huguet2018earth, wilson2023can} and growth \citep{walker2025non}, quantitative predictions are consistent with the primary observations that constrain the F-layer. Specifically, 1D steady solutions of the non-rotating and non-magnetic equilibrium two-phase two-component slurry equations produce stably stratified F-layers over a wide range of parameters that span plausible values for Earth's core and are consistent with the ICB density jump and compressional wave speeds observed by seismology \citep{wong2021regime}. 

In this paper, we seek geomagnetic evidence for the existence of the slurry F-layer. Available observations do not directly probe the magnetic field interior to the core, and so we make use of global time-dependent reconstructions of the field spanning historical \citep{jackson2000four}, Holocene \citep{cals10k2}, and multi-millennial \citep{panovska2019one} timescales that can be downward continued to the core-mantle boundary (CMB). Despite the fact that the F-layer is thousands of kilometres below the CMB, indirect evidence from dynamo simulations nevertheless suggests that it could influence the observable field. \citet{stanley2006numerical} modelled dynamo action in a thin layer underlain by a stably stratified region and found that this configuration could explain the non-axisymmetric non-dipolar fields of Uranus and Neptune, which contrasts with the dipole-dominated fields that are usually produced in thick-shell dynamos without stable regions \citep[e.g.][]{christensen2006scaling, tassin2021geomagnetic}. More recently, \citet{landeau2017signature} found that changing the inner core size and core buoyancy distribution altered the latitudinal variation of CMB field morphology, which resulted in part from the changing location of the imaginary tangent cylinder (TC) that circumscribes the inner core equator. Quasi-geostrophic core dynamics induce columnar fluid structures in the direction parallel to the rotation axis that would have to significantly change their length to pass the TC, which is unlikely under rotationally constrained conditions, leading to distinct dynamics inside and outside the TC \citep[e.g.][]{j2015}. We might therefore expect a similar effect if the presence of the F-layer acts as a barrier to columnar flow. We, therefore, hypothesise that the presence of the F-layer could induce latitudinal variations in the CMB magnetic field. 

We conduct non-magnetic rotating spherical shell convection and dynamo simulations in a configuration that consists of a thermally unstable and chemically stable F-layer (with overall stability against convection) underlying a convecting outermost core. Owing to the significant numerical challenge of directly simulating the two-phase two-component slurry equations \citep[which comprise 13 coupled PDEs and non-linear boundary conditions;][]{wilczynski2025two}, we make a number of simplifications to the F-layer dynamics. The 1D steady solutions to the full slurry equations found by \citet{wilczynski2025two} at geophysically relevant conditions revealed the following characteristics: liquid velocity and solid fraction are very small, resulting in an F-layer predominantly comprised of stagnant liquid; thermal gradients are destabilising with latent heat released at the ICB approximately equal to the heat extracted from the F-layer (i.e., there is no source of heat generation in the layer); compositional gradients are stabilising, with light material transported to the top of the F-layer by barodiffusion and advection. These results suggest that a simple representation of the F-layer requires separate thermal and chemical fields that combine to yield a strong net stratification. For simplicity, we ignore the presence of any solid phase and consider that the internal dynamics of the F-layer arise due to perturbations caused by interactions with the overlying liquid core. This allows us to implement the F-layer in the steady 1D reference state of the Boussinesq equations, as is done in studies of stable stratification in the upper regions of the core \citep[e.g.][]{nakagawa2011effect, christensen2018geodynamo, gastine2020dynamo}. Our primary focus is to understand if and how the F-layer influences the flow and field morphology at the top of the core. 

This paper is organised as follows. Section~\ref{sec:methods} contains a description of the standard Boussinesq geodynamo equations together with details of the implementation of the F-layer. We also define the dimensionless parameters that control the thermochemical convection and dynamo, as well as the main diagnostic parameters used later. First, we present the results regarding the effect of the presence of an F-layer in thermochemical simulations (see Section~\ref{sec:results_nonmag}). Then, we present dynamo simulations with an F-layer, and we compare the magnetic field produced to recent or paleo-observations of the geomagnetic field (Sec~\ref{sec:results_mag}). In section~\ref{sec:discussion}, we discuss the implications of our results to constrain the strength or thickness of the F-layer in the Earth's core.

\section{Problem formulation}
\label{sec:methods}

\subsection{Model equations}

We simulate Boussinesq thermochemical convection and dynamo action in a spherical shell rotating about the vertical $\zhat$ direction at angular frequency $\Omega$. The inner and outer boundaries and the initial F-layer radius are respectively denoted $\ri$, $\ro$ and $\rs$ in the spherical polar coordinate system $(r, \theta, \phi)$ and gravity varies linearly with radius as $\mathbf{g} = -g_o (r/\ro)\mathbf{\hat{r}}$ where $g_o$ is the acceleration of gravity at $\ro$ and $\mathbf{\hat{r}}$ is the unit vector in the radial direction. We assume constant values of the core's density $\rho$, kinematic viscosity $\nu$, coefficient of thermal expansion $\alpha_T$, coefficient of compositional expansion $\alpha_C$, magnetic diffusivity $\eta$, magnetic permeability $\mu_0$, thermal diffusivity $\kappa_T$, and chemical diffusivity $\kappa_C$. We solve for the evolution of the temperature $T$, composition $C$, velocity $\vel$, and magnetic field $\magf$. We use $\prime$ to indicate radial derivatives, such that $\dTsdr$ is the temperature gradient at $\ri$ and $\dcsdr$ the compositional gradient at $\rs$. Scaling length by $D = \ro - \ri$, time by $D^2/\eta$, magnetic field by $(2\rho\mu_0 \eta \Omega)^{1/2}$, temperature by $\dTsdr D$, and light element concentration by $\dcsdr D$, the dimensionless governing equations are then:
\begin{align}
\frac{\partial \vel }{\partial t}
&= \mathrm{Pm}\,\nabla^{2} \vel - \nabla\left(\frac{\mathrm{Pm}}{\Ek}\,\tilde{P} + \frac{1}{2}\lvert \vel \rvert^{2}\right) + \vel \times (\nabla \times \vel) \nonumber\\
&\quad - \frac{2\mathrm{Pm}}{\Ek}\,\zhat \times \vel
+ \mathrm{Pm}^{2}\left(\frac{\mathrm{Ra}_{T}}{\mathrm{Pr}_{T}}\,T + \frac{\mathrm{Ra}_{C}}{\mathrm{Pr}_{C}}\,C\right)\, \mathbf{r}
+ \frac{2\mathrm{Pm}}{\Ek}\,(\nabla \times \magf) \times \magf
\label{eq:momentum}
\end{align}

\begin{equation}
\frac{\partial \magf}{\partial t} = \nabla \times (\vel \times \magf) + \nabla^{2} \magf,
\end{equation}

\begin{equation}
\frac{\mathrm{Pr}_{T}}{\mathrm{Pm}}\left(\frac{\partial T}{\partial t} + \vel \cdot \nabla T\right) = \nabla^{2} T,
\end{equation}

\begin{equation}
\frac{\mathrm{Pr}_{C}}{\mathrm{Pm}}\left(\frac{\partial C}{\partial t} + \vel \cdot \nabla C\right) = \nabla^{2} C + S,
\end{equation}

\begin{equation}
    \nabla \cdot \vel = \nabla \cdot \magf = 0,
\end{equation}
where $\tilde{P}$ is the modified pressure, and $S$ denotes the compositional source term. In the governing equations, the dimensionless parameters are
$$ \mathrm{Ra}_{T} = \frac{\alpha_T g_o \dTsdr D^5}{\nu \kappa_T \ro}, \quad \mathrm{Ra}_{C} = \frac{\alpha_C g_o C_s^{\prime} D^5}{\nu \kappa_C \ro}, \quad \mathrm{Pr}_{T} = \frac{\nu}{\kappa_T}, \quad \mathrm{Pr}_{C} = \frac{\nu}{\kappa_C}, \quad \Ek= \frac{\nu}{\Omega D^2}, \quad \Pm=\frac{\nu}{\eta} .$$

From here on, we will refer to dimensionless quantities. The reference state of the thermal component, with fixed flux at both boundaries, gives $\partial T/\partial r = -1/r^2$. We implement the F-layer within the 1D steady reference state described by $\vel = \magf = 0$. We define the radius $\rs$ as the radius where the reference state concentration gradient takes a value $\dcsdr$, which is negative and hence destabilising. We also define the width of the stratified region in the reference state $H_\textrm{S}=r_s-r_i$. The actual thickness of the F-layer $\hf$ is smaller than $H_\textrm{S}$ because a ``mixing layer'' exists to transition between the stratified and convecting regions \citep{wong2018boussinesq}. The region between $\ri$ and $\rs$ will be denoted by subscript $L$ (for lower), while the region between $\rs$ and $\ro$ will be denoted by subscript $U$ (for upper).

The compositional source term $S$ differs between the F-layer and the layer above it, $ S_L$ and $ S_U$, respectively, to impose the compositional gradient $\dcsdr$ and mass conservation within and above the F-layer. The reference state is defined by
\begin{equation}
    \frac{1}{r^2} \frac{d}{dr} \left(r^2 \dfrac{d C}{dr}\right) = \begin{cases} S_U & \text{if } r\ge r_s, \\S_L & \text{if } r< r_s.\end{cases}
    \label{eq:refstate_comp}
\end{equation}
and by the boundary conditions 
\begin{align}
   \left. \frac{dC}{dr} \right|_{r=\ro} &=0,\qquad \text{and} \qquad     \left. \frac{dC}{dr} \right|_{r=\ri}=\dcidr.\label{eq:dcrodcri}
\end{align}
From Eq.~\ref{eq:refstate_comp}, the source terms are defined as
\begin{align}
S_U  = -\frac{3\rs^2 \dcsdr}{ (\ro^3 - \rs^3)} \qquad \text{and} \qquad S_L  = -\frac{3(\dcidr \ri^2 - \dcsdr\rs^2 )}{(\rs^3 - \ri^3)}, \label{eq:susl}
\end{align}
with 
\begin{equation}
    \left. \frac{dC}{dr} \right|_{r=\rs} = \dcsdr= -\frac{\rs-\frac{\ro^3}{\rs^2} }{\ro^3 - \ri^3}. \label{eq:dcrs}
\end{equation}

To complete the problem specification, we impose no-slip conditions on both boundaries, fixed thermal flux on $\ri$ and $\ro$, and fixed chemical flux as described in Eqs.~(\ref{eq:dcrodcri}-\ref{eq:dcrs}). For dynamo simulations, we used insulating conditions at the outer core boundary and either insulating (DI) or conductive (DC) conditions at the inner core boundary.

The stabilising compositional gradient at $\ri$ in our Boussinesq simulations is not constrained by observations. To set its value, we use the dimensionless Brunt-Väisälä frequency, which can be written
\begin{equation}
\mathcal{N}^2=\frac{N^2}{\Omega^2} = r \Ek^2\left(\frac{\mathrm{Ra}_{T}}{\mathrm{Pr}_{T}} \frac{\partial T}{\partial r}+\frac{\mathrm{Ra}_{C}}{\mathrm{Pr}_{C}} \frac{\partial C}{\partial r}\right) \equiv \frac{N_T^2}{\Omega^2} + \frac{N_C^2}{\Omega^2}.
\end{equation}
$\dcidr$ can be written as a function of $\mathcal{N}^2_\textrm{i}$ (at the inner core boundary):
\begin{equation}
\dcidr= \left[\frac{\mathcal{N}^2_\textrm{i}}{\ri \Ek^2} + \left(\frac{\mathrm{Ra}_{T}}{\mathrm{Pr}_{T}} \frac{1}{\ri^2}\right) \right] \frac{\mathrm{Pr}_{C}}{\mathrm{Ra}_{C}} .
\label{eq:BVadim}
\end{equation}
Therefore, with our parameter selection (see Sec. \ref{sec:param}) and values of $\dcidr$ of 5, 10, 20, 80 and, 100, we investigate cases with weak stratification $\mathcal{N}^2_\textrm{i} < 1$ and cases with strong stratification $\mathcal{N}^2_\textrm{i}>1$. We call $\hf=r(\mathcal{N}^2=0)-r_\textrm{i}$ the actual width of the stratified region in the simulations using the radial profile of $\mathcal{N}^2(r)$ averaged with time, azimuth and longitude.

\subsection{Parameter Selection}
\label{sec:param}

The geodynamo problem with a slurry F-layer is governed by 8 dimensionless parameters: $\ri/\ro$, $\Ek$, $\mathrm{Pr}_{T}$, $\mathrm{Pr}_{C}$, $\mathrm{Ra}_{T}$, $\mathrm{Ra}_{C}$, $\rs$, $\dcidr$ and $\dcsdr$. 
To reduce this space to a tractable number of calculations, we fix several of the parameters that control the bulk core dynamics and focus on varying parameters that dictate the behaviour of the F-layer. To this end, we set $\ri/\ro=0.35$ in the majority of our simulations, which is the aspect ratio for Earth's modern core. We also test in 1 simulation the effect of increasing $\ri/\ro$ to mimic a thicker F-layer. We use $\mathrm{Pr}_{T}=1$, $\mathrm{Pr}_{C}=10$, and $\Ek=2\times 10^{-5}$, which were employed in the work of \cite{nmdc2025} and are similar to values used in recent simulations of top-heavy double diffusive convection \citep{tassin2021geomagnetic}. A value of $\mathrm{Pr}_{T}=1$ is frequently adopted in geodynamo simulations, which is somewhat larger than the estimated value for Earth’s core, $\mathrm{Pr}_{T}\sim 0.05$ \citep{pozzo2013transport}. The chosen $\mathrm{Pr}_{C}$ is smaller than the core estimate $\mathrm{Pr}_{C}\sim 100$ \citep{pozzo2013transport} and is selected primarily for computational convenience. The ratio $\mathrm{Pr}_{C}/\mathrm{Pr}_{T}$ influences the scale separation between thermal and chemical fields \citep{calkins2012influence} and hence this effect is underestimated in our simulations compared to Earth's core. The value of $\Ek$ is set low enough that the leading-order force balance is quasi-geostrophic and viscosity is subdominant to all terms \citep{naskar2025force}, but high enough to allow a reasonable number of simulations to be run until a statistically steady state is realised. 

The thermal and chemical Rayleigh numbers, $\mathrm{Ra}_{T}$ and $\mathrm{Ra}_{C}$ respectively, are set based on physical considerations and the value of $\Ek$. In Earth's core, $\mathrm{Ra}_{C} \gg \mathrm{Ra}_{T}$ \citep{gubbins2001rayleigh, jones2000convection} and the convective power supplied by compositional buoyancy (which is $\propto \mathrm{Ra}_{C}/\mathrm{Pr}_{C}$) strongly exceeds the power supplied from thermal buoyancy ($\propto \mathrm{Ra}_{T}/\mathrm{Pr}_{T}$) \citep{lister1995strength, davies2015cooling, nimmo2015energetics}. Furthermore, in the F-layer, the stabilising chemical buoyancy must strongly exceed the destabilising thermal buoyancy to produce net stratification \citep{gubbins2008thermochemical, wilczynski2025two}. In our simulation setup, $\mathrm{Ra}_{T}$ determines the amplitude of thermal buoyancy everywhere in the core; we therefore require $\mathrm{Ra}_{C}\gg \mathrm{Ra}_{T}$ to obtain solutions relevant to the geophysical case in which compositional convection is believed to dominate \citep{davies2015constraints,labrosse2015thermal}. At the chosen $\Ek$, the critical Rayleigh number for pure chemical convection with $\mathrm{Pr}_{C}=10$ and for pure thermal convection with $\mathrm{Pr}_T=1$ are $34.1\times 10^{5}$ and $24.7\times 10^{5}$, respectively \citep{nmdc2025}. We therefore set $\mathrm{Ra}_{C}=10^9$, which is sufficiently above the critical value that the non-magnetic system is vigorously convecting but not so large that the constraint of rotation is lost \citep{nmdc2025}. For dynamo simulations, this value produces strong-field ($\magEnergy > \kinEnergy$) dipole-dominated solutions that do not reverse polarity. We consider three values of $\mathrm{Ra}_{T}$, $90\times10^{5},\ 55\times10^{6},\ 12\times10^{7}$, to assess its impact on the solutions. 

The remaining parameters, $\dcsdr$, $\dcidr$, and $\rs$, describe the behaviour of the F-layer. To facilitate comparison with previous results, we set $\dcsdr$ to the reference state value at $\rs$ that exists in the absence of the F-layer. Changing the value of $\dcsdr$ is equivalent to changing $\mathrm{Ra}_{C}$, and so the value we have chosen ensures that the simulations maintain plausible dynamics as described above. The value of $\dcidr$ is not well constrained, and so here we set its value using $\mathcal{N}^2$, which is $\sim$1 for the strong stabilising density gradients inferred from the seismic observations \citep{gubbins2008thermochemical}. To vary the degree of F-layer stratification, we therefore vary $\dcidr$ to obtain a given $\mathcal{N}^2_\textrm{i} \ = 0.01-2$. For $\rs$, we consider values $\rs = 0.63, 0.83, 1.03$, which correspond in dimensional unit to $226$, $679$, and $1132$~km above the inner boundary. The latter two values are larger than the F-layer thicknesses obtained from seismic observations. We use these values because initial calculations showed that the convection can significantly reduce the thickness of the stable region $\hf$, compared to the thickness imposed in the reference state by $\hs$. 

To summarize, we vary $\dcidr$, $\mathrm{Ra}_{T}$, and $\rs$ with $\dcidr \in \left[5,10,20,80,100\right]$, $\mathrm{Ra}_{T}\in \left[90 \times 10^5,\, 55 \times 10^6,\, 12 \times 10^7\right]$ and $\rs\in \left[0.63,0.83,1.03\right]$ (corresponding to a distance above the inner core boundary of $\hs=r_s-r_i\in \left[0.1,0.3,0.5\right]$).
Other parameters are fixed at $\Ek=2\times10^{-5}$, $\mathrm{Pr}_{T} = 1$, $\mathrm{Pr}_{C} = 10$, and $\mathrm{Ra}_{C} = 10^9$. We have performed 34 thermochemical convection (named CXX, C01 to C04 for no-F-layer thermochemical simulations, C05-C34 for F-layer thermochemical simulations). We have also performed 10 dynamo simulations, named DIXX or DCXX depending on the insulating or conductive boundary conditions at the ICB. DX01 to DX03 are no F-layer dynamo simulations, and DX04 to DX10 are F-layer dynamo simulations. The input parameters for each simulation can be found in Table 1 of the supplementary dataset \citep{hfnmcrmd2026}.

\section{Results}
\label{sec:results}

In this section, we present an analysis of the influence of a stably stratified F-layer on rotating spherical shell convection and dynamo action. We first consider non-magnetic rotating convection in section~\ref{sec:results_nonmag}, which helps to develop hypotheses for the influence of the F-layer in the dynamo simulations analysed in section~\ref{sec:results_mag}. 

\subsection{Non-magnetic Convection Simulations}
\label{sec:results_nonmag}

\begin{figure}
    \centering
    \hspace{5.5mm}
    \includegraphics[trim={1cm 8.0cm 15cm 0cm},clip]{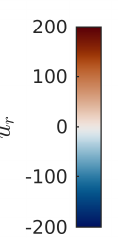}
    \begin{overpic}[width=0.18\linewidth,trim={2.4cm 0cm 0cm 0cm},clip]
        {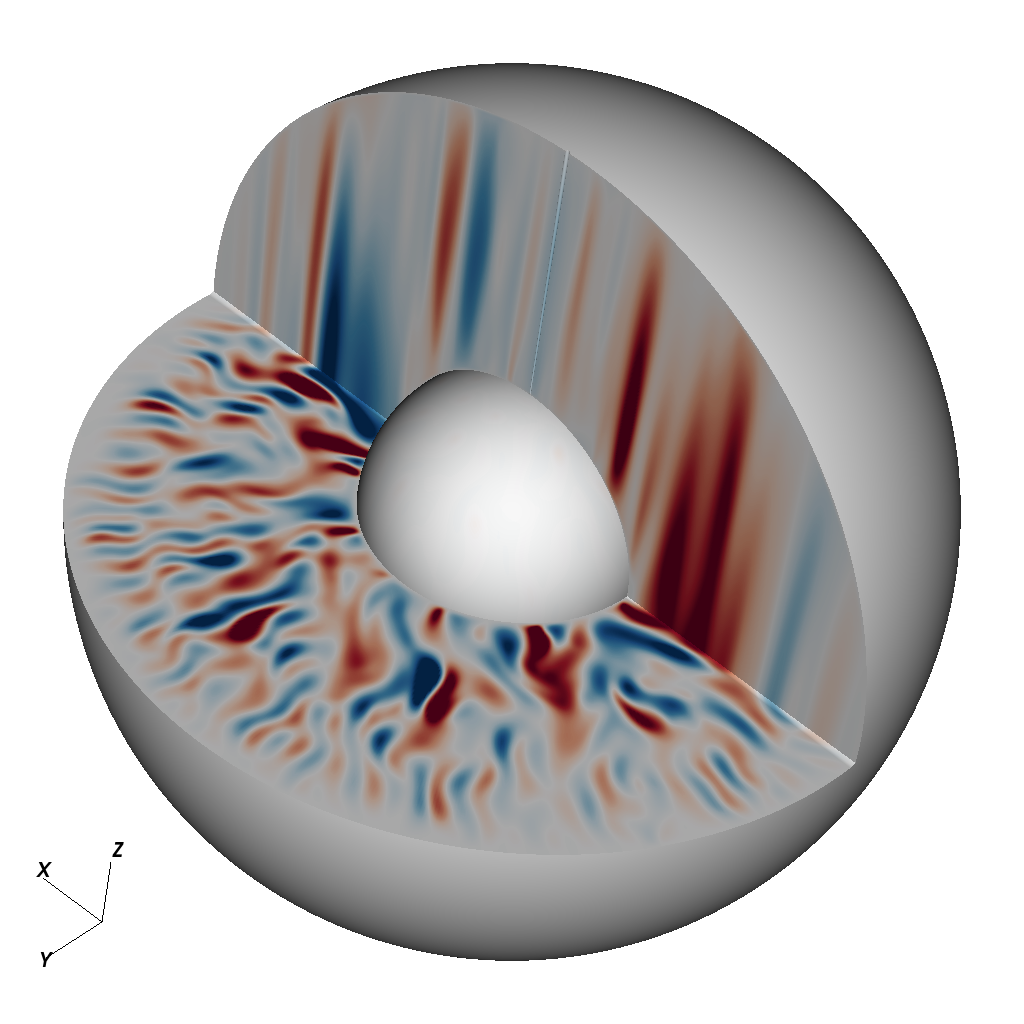}
        \put(10,100){$\textrm{case}\ \textrm{C02}$}
    \end{overpic}
    \hspace{2mm}
    \begin{overpic}[width=0.18\linewidth,trim={2.4cm 0cm 0cm 0cm},clip]
        {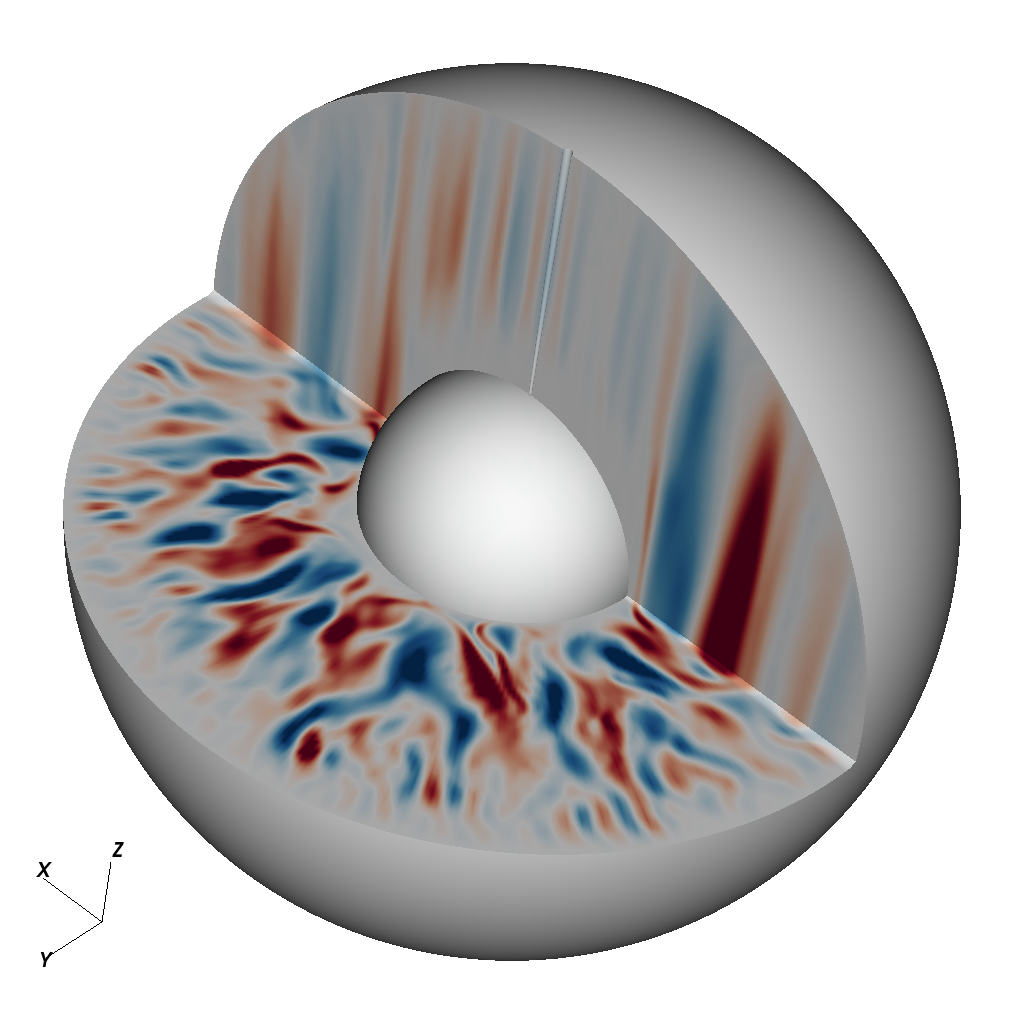}
        \put(10,100){$\textrm{case}\ \textrm{C18}$}
    \end{overpic} 
    \hspace{2mm}
    \begin{overpic}[width=0.18\linewidth,trim={2.4cm 0cm 0cm 0cm},clip]
        {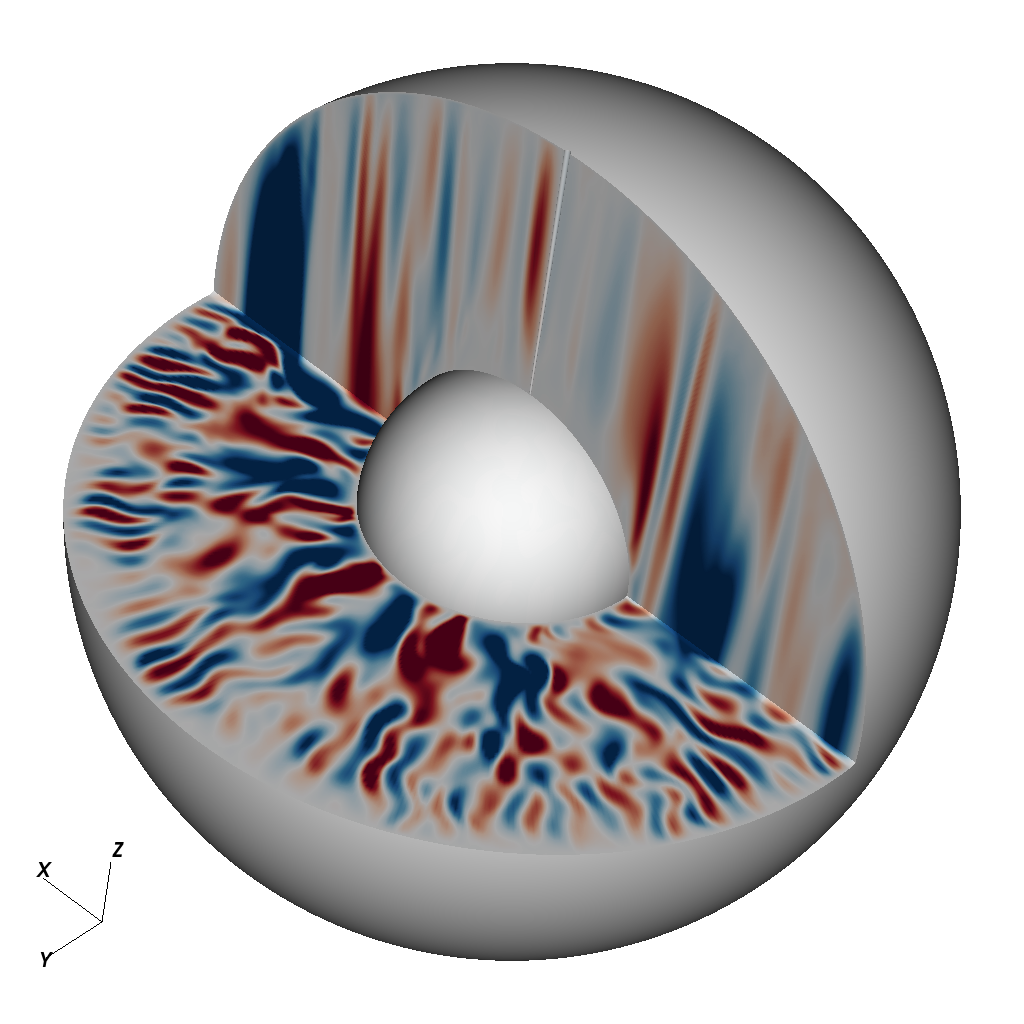}
        \put(10,100){$\textrm{case}\ \textrm{C19}$}
    \end{overpic}
    \hspace{2mm}
    \begin{overpic}[width=0.18\linewidth,trim={2.4cm 0cm 0cm 0cm},clip]
        {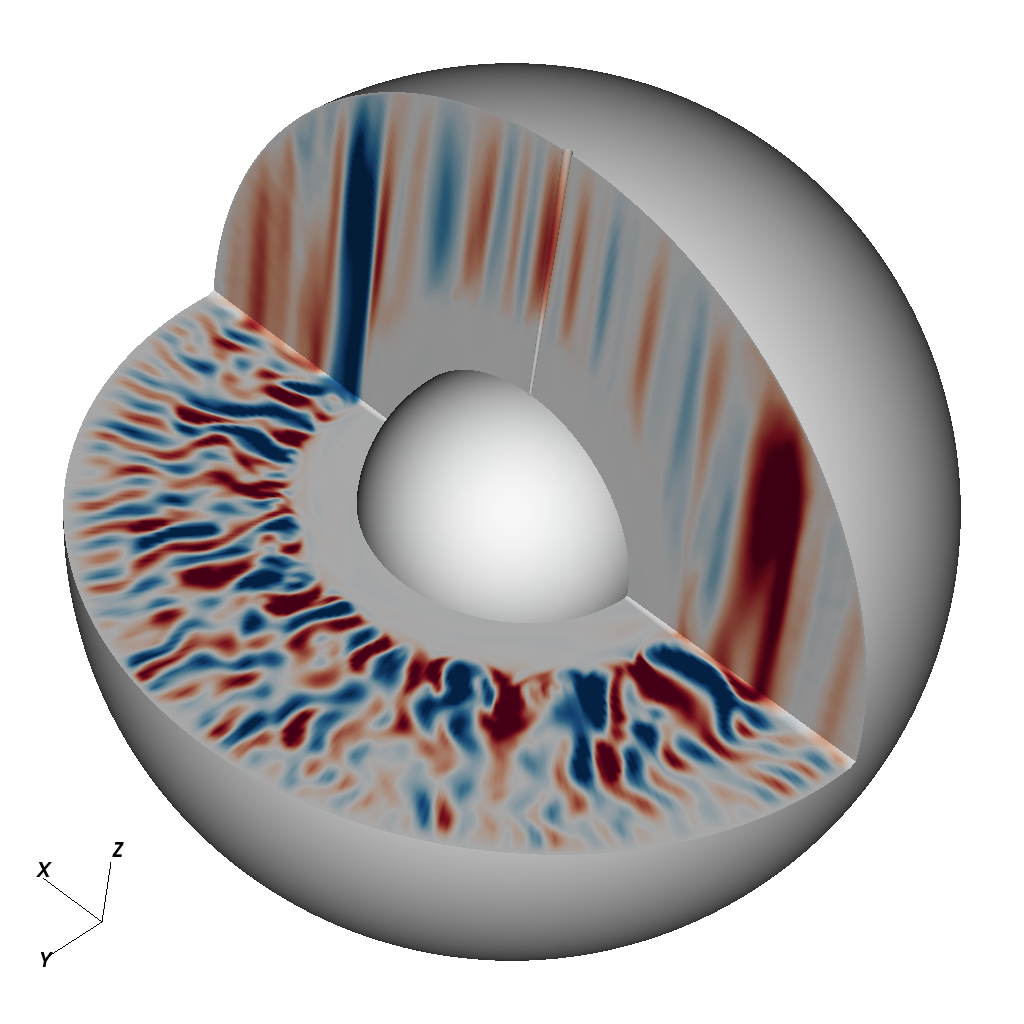}
        \put(10,100){$\textrm{case}\ \textrm{C34}$}
    \end{overpic} 
    \includegraphics[trim={0cm 0cm 0cm 4cm},clip]{FIGURES/figure_ur_uph_BV_3D_2D.pdf}
    \caption{Pseudo isometric plot of instantaneous $u_r$ (top row),  time and azimuthal average of $u_\phi$ (middle row), and time and azimuthal average of $\mathcal{N}^2$ (bottom row). Columns show four simulations: C02 ($\mathrm{Ra}_{T} = 55 \times 10^6;\, \dcidr=-0.4725$), C18 ($\mathrm{Ra}_{T} = 55 \times 10^6;\, \dcidr=10$),  C19 ($\mathrm{Ra}_{T} = 12 \times 10^7;\, \dcidr=10$), C34  ($\mathrm{Ra}_{T} = 12 \times 10^7;\, \dcidr=100$). In the bottom row, the blue line denotes the iso-contour for $\partial C/\partial r= 0$, and the light blue area denotes the stable region with $\mathcal{N}^2>0$. Vertical dotted lines indicate the tangent cylinder of the inner core. Black dashed lines show the radius for which $\mathcal{N}^2= 0$ when averaged in time, $\phi$ and $\theta$, white dashed lines correspond to the radius $\rs$ in the reference state.}
    \label{fig:nonmag_merid}
\end{figure}

Fig.~\ref{fig:nonmag_merid} shows isometric projections of instantaneous radial velocity $u_r$, and time- and azimuthally-averaged meridional sections of azimuthal velocity $u_{\phi}$ and normalized squared buoyancy frequency $\mathcal{N}^2$ for case C02 without an F-layer and cases C18, C19, C34 with $r_s=0.83$ ($\hs=0.3$) and varying F-layer properties. Case C02 shows well-known features: unstable stratification at all radii ($\mathcal{N}^2<0$), columnar convective flow structures of small horizontal length scale aligned with the rotation axis, and a dominantly zonal time-averaged flow that is retrograde outside the tangent cylinder. Case C18 has $\mathrm{Ra}_{T} \ll \mathrm{Ra}_{C}$, as expected for the outer core \citep{gubbins2001rayleigh}, and a relatively weak chemically stabilising gradient imposed at $\ri$ that is nevertheless strong enough to produce a global stratification within the F-layer. Deep within the F-layer, where the chemically-induced stability is strongest, the stratification is only weakly perturbed by the thermal field. However, the F-layer thickness calculated from the mean of the $\mathcal{N}^2=0$ contour is significantly less than that imposed via the reference state (i.e., $\rs$), indicating that thermally destabilising effects dominate near the top of the layer. Radial motion is suppressed within the layer, whereas zonal flows are present within the layer since these are not directly influenced by the stratification. In case C19, $\mathrm{Ra}_{T}$ is nearly double the value in case C18, reducing the F-layer thickness by promoting thermally destabilising buoyancy and increasing the amplitude of $u_r$ and $u_{\phi}$. In case C34, the amplitude of the chemically stabilising gradient is increased by a factor of 10 from that in case C19, which increases the F-layer thickness by a factor of 4 (from 165 km to 626 km). It also increases the radially-averaged strength of the stratification by a factor of 1.5 (from $<\mathcal{N}> = 0.56$ to 0.86, for C19 and C34, respectively), which suppresses the radial motion throughout the layer ($<>$ denote the radial averaged in the stable layer). In case C34 (and in case C18), the dominant axially aligned zonal flow structure is moved away from the inner boundary such that it sits just outside $\rs$. In all cases, the scale of flow and thermochemical anomalies in the bulk is relatively unaffected by the presence of the F-layer.

In the bottom row of Fig.~\ref{fig:nonmag_merid}, we further note that the thickness of the stably stratified layer varies with latitude (longitudinal variations are smaller), being greater than the mean at the pole and smaller at the equator. The corresponding thickness variations are approximately 1\%, 5\%, and 11\% for $\mathrm{Ra}_T =\left[90 \times 10^5,\, 55 \times 10^6,\, 12 \times 10^7\right]$, respectively. The stratification strength is generally anti-correlated with the layer thickness, exhibiting latitudinal variations of about 5\%, 20\%, and 30\% as $\mathrm{Ra}_T$ increases. Moreover, for very strong stratification, $\mathcal{N}^2_\textrm{i}\sim 2$ (C34), both the thickness and stratification variations become even smaller.

\begin{figure}[!h]
    \centering
    \includegraphics[width=1\textwidth]{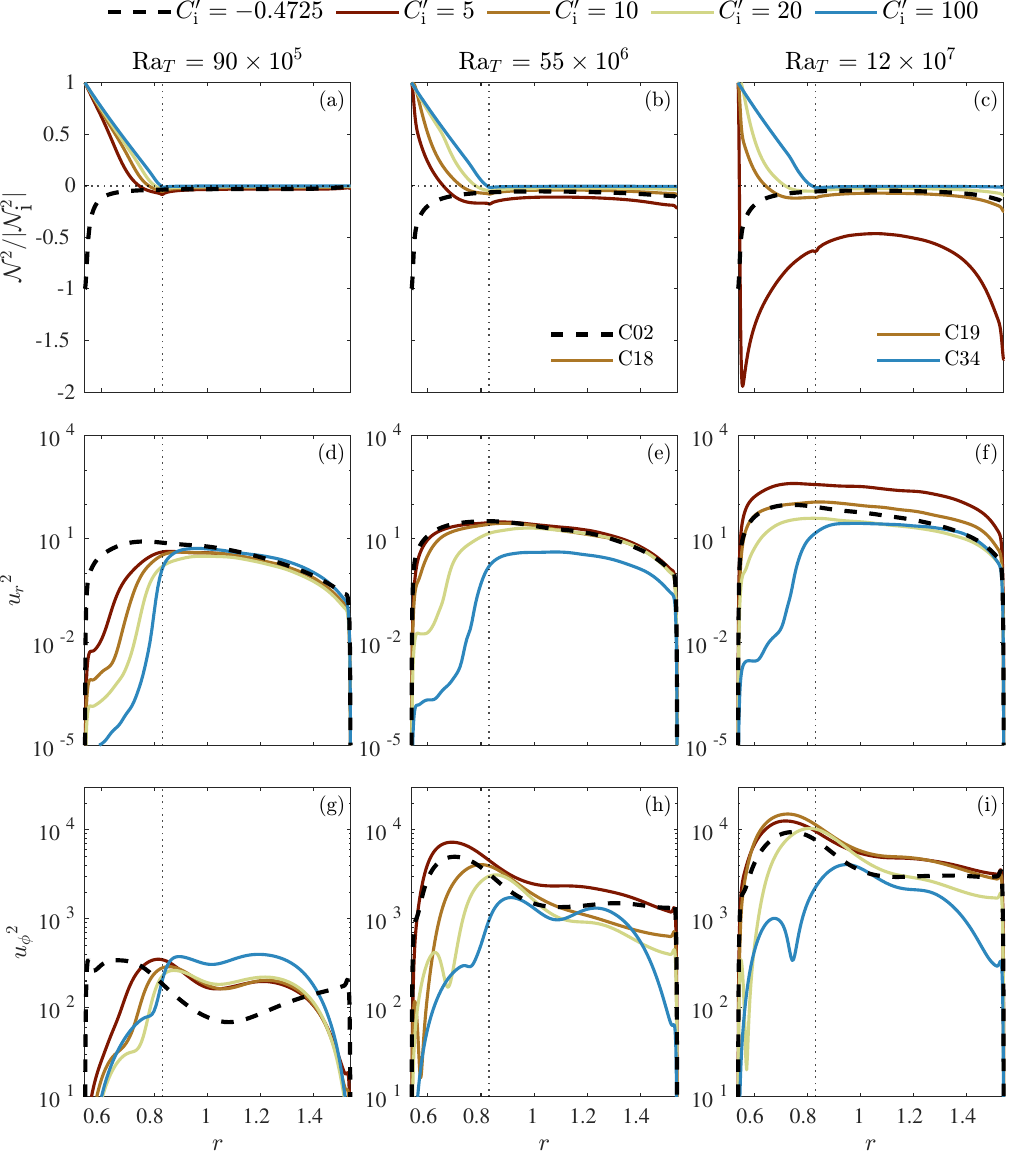}
    \caption{Radial profiles of the dimensionless Brunt-Väisälä ($\mathcal{N}^2$) (a, b, c),  of the square of radial velocity ($u_r^2$) (d, e, f), and of the square of azimuthal velocity ($u_\phi^2$) (g, h, i). Each colour denotes a different value of $\dcidr$. Columns from left to right correspond to $\mathrm{Ra}_{T}=90 \times 10^5$, $55 \times 10^6$ and $12 \times 10^7$, respectively.  For comparison, $\mathcal{N}^2$ is normalised by its absolute value at the inner core boundary (where it is maximum and set by the imposed by $\dcidr$). Vertical dotted lines denote the radius $\rs$ in the reference state. In the middle and right columns, 4 simulations correspond to the ones shown in Fig.\ref{fig:nonmag_merid}.}
    \label{fig:nonmag_radial}
\end{figure}

Fig.~\ref{fig:nonmag_radial} shows the radial variations of $u_r^2$, $u_{\phi}^2$ and $\mathcal{N}^2$. It is clear that increasing $\dcidr$ increases the radius of the stable region towards the reference value $\rs$, decreases the amplitude of the destabilising chemical gradient at $\rs$, and decreases $u_r$ and $u_{\phi}$ in the F-layer. The trends for layer thickness and $u_r$ are essentially reversed with increasing $\mathrm{Ra}_{T}$, which also causes a substantial increase in $u_{\phi}$ within the F-layer. However, the values of $\mathrm{Ra}_{T}/\mathrm{Ra}_{C}$ adopted in our study (i.e., $\mathrm{Ra}_{T}/\mathrm{Ra}_{C} \sim 0.009$, $0.05$, and $0.12$) are likely to be much larger than the geophysically relevant values with $\mathrm{Ra}_{T}/\mathrm{Ra}_{C} \sim \mathcal{O}(10^{-10})$ in Earth's core \citep{gubbins2001rayleigh}. This implies that the thermally destabilising influence is probably overestimated in our simulations. A related point is that the reduction in $u_r$ in the F-layer compared to the bulk core is less than 1 order of magnitude for all simulations with $\mathrm{Ra}_T=55\times 10^6$ and $12\times 10^7$, which is much weaker than the reduction predicted by the simulations of \citet{wilczynski2025two}. 

\begin{figure}
    \centering
    \includegraphics[width=1\textwidth]{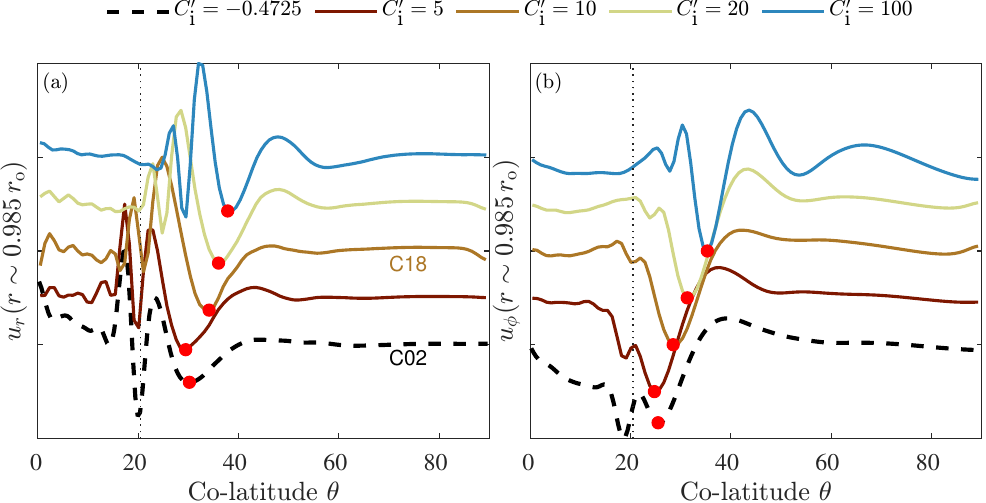}
    \caption{Profiles of $u_r$ (a) and $u_\phi$ (b) as a function of co-latitude at a radius just below the outer core boundary (i.e., below the boundary layer which is defined by a local maximum of horizontal velocity $\sqrt{\vec{u}_h^2}$, near the core mantle boundary). Plotted velocities are normalised by their respective maximum value, and for different values of $\dcidr$, a vertical offset is applied to each case to introduce visual separation of the profiles. For these thermochemical simulations, $\mathrm{Ra}_{T}$ is $55 \times 10^6$. Vertical dotted lines denote the tangent cylinder. Red dots indicate the position of the minimum of $u_r$ (a), which corresponds to a down-welling outside the TC, and the position of the minimum of $u_\phi$ (b), which corresponds to westward flow outside the TC. Dashed black and solid brown lines correspond to simulations C02 and C18 seen in Fig.\ref{fig:nonmag_merid}). }
    \label{fig:nonmag_lat}
\end{figure}

Fig.~\ref{fig:nonmag_lat} shows $u_r(\theta)$ and $u_{\phi}(\theta)$ at a radius below the upper boundary layer (defined by a local maximum of horizontal velocity $\sqrt{\vec{u}_h^2}$, near the core mantle boundary; recall that $\mathbf{u} = 0$ on $r=\ro$) for simulations with $\mathrm{Ra}_{T} = 55\times10^6$. In the absence of an F-layer, there are minima in $u_{\phi}$ and $u_r$ just outside the tangent cylinder. The former is associated with the zonal flow (Fig.~\ref{fig:nonmag_merid}) while the latter arises where high-latitude and low-latitude meridional circulations converge. Increasing the thickness and stratification strength of the F-layer moves these minima to lower latitudes. Since core surface downwellings are expected to concentrate the magnetic field \citep[e.g.][]{gubbins2003thermal}, we might expect that this feature (if it persists in the dynamo simulations) will present a signature in $B_r$ at $\ro$. In particular, we expect peaks in $B_r(\ro,\theta)$ to arise at lower latitudes as $\rs$ and $\dcidr$ are increased.

\begin{figure}
    \centering
    \includegraphics[width=1\textwidth]{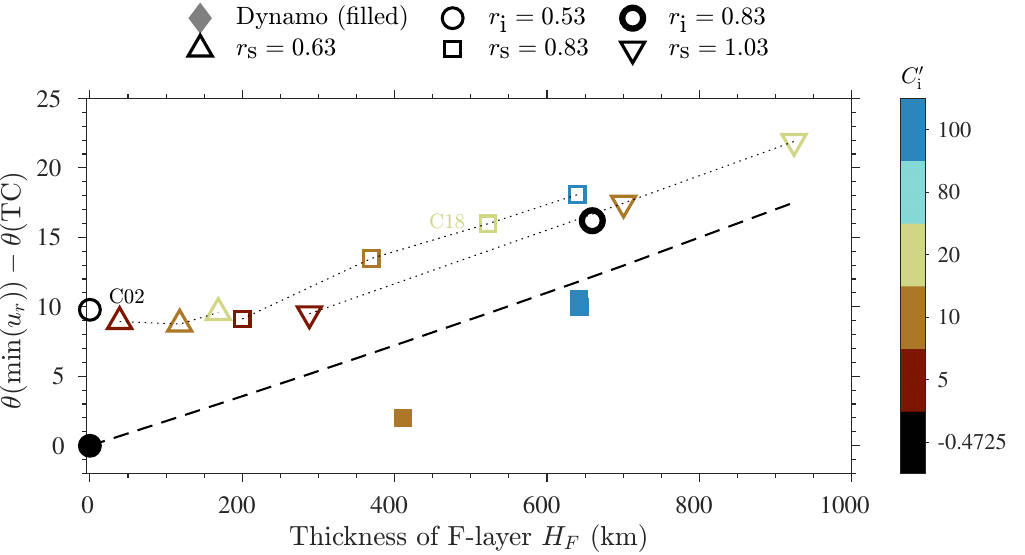}
\caption{Co-latitudinal distance between the minimum of $u_r$ and the tangent cylinder of the inner core as a function of the thickness of the stable F-layer for simulations with $\mathrm{Ra}_{T}= 55 \times 10^6$. Filled and empty circles denote dynamo and thermochemical simulations, respectively. Triangles, squares, and upside-down triangles denote source radius $r_s= 0.63,\, 0.83,\, 1.03$, respectively. Thin and thick circles denote no-F-layer simulations (with $r_s= 0$) for two different inner core radii ($\ri=0.538\textrm{ or }\, 0.83$), respectively. Six colours show the different values of the strength of the stratification $\dcidr$. The black dashed line denotes the position of the tangent cylinder of the F-layer relative to the inner core tangent cylinder. Dotted lines highlight the effect of varying $\dcidr$ with all other parameters fixed. }
\label{fig:max_ur}
\end{figure}

Fig.~\ref{fig:max_ur} shows the co-latitudinal distance between the position of a minima in $u_r$ (as identified in Fig~\ref{fig:nonmag_lat}(a)) and the tangent cylinder, as a function of the actual F-layer thickness $\hf$ (as measured by the $\mathcal{N}^2=0$ contour) for all simulations with $\mathrm{Ra}_{T}=55 \times 10^6$. Increasing $\dcidr$ or $\rs$ increases the F-layer thickness and moves the minimum of $u_r$ to lower latitudes. Dynamo simulations have thinner F-layers, and hence a higher latitude of minimum of $u_r$, compared to the non-magnetic runs at equivalent parameters. Fig.~\ref{fig:max_ur} also highlights trade-offs between $\rs$ and $\dcidr$. For example, similar F-layer thickness and minimum $u_r$ latitude are obtained for $[\dcidr=100, \rs=0.83]$ and $[\dcidr=10, \rs=1.03]$. A weakly stratified layer (for $\rs = 0.63$ or $\dcidr\le5$) does not change the position of the strong downwelling near the tangent cylinder compared to the thermochemical case without F-layer (C02). 

Fig.~\ref{fig:max_ur} also shows an additional non-magnetic simulation with a larger inner core ($\ri = 0.83$) to test whether this change alone could mimic the behaviour of the stratified F-layer. This simulation produced a minimum in $u_r$ at $16^{\circ}$ equatorward of the TC (of the present-day inner core), a similar latitude to the F-layer calculation with $\dcidr=100$ and $\rs=0.83$ or $\dcidr=10$ and $\rs=1.03$. However, these two F-layer simulations have a different stratified layer with $\mathcal{N}_\textrm{i}^2=2.11$ ($<\mathcal{N}^2>=0.87$) and $\mathcal{N}_\textrm{i}^2=0.17$ ($<\mathcal{N}^2>=0.076$), while having about the same thickness ($\sim 650$~km). Even if a large inner core can reproduce some of the behaviour of the F-layer, the strength of the stratification plays a role in controlling how columns can penetrate the tangent cylinder. 
We therefore conclude that simply increasing the size of the inner core does not adequately capture the influence of the stratified F-layer. 

The behaviour in Figs.~\ref{fig:nonmag_lat} and \ref{fig:max_ur} arises from the quasi-geostrophic dynamics that predominate in the simulations. Analysis of force spectra reveals that, in all cases (e.g. see Figure~\ref{SI:fspec}), the leading-order balance at large scales involves the Coriolis force and the pressure gradient. The ageostrophic part of the Coriolis force is balanced at the next order (and at large scales) by the buoyancy force and inertia, with viscosity playing a subdominant role. In this regime, the convection takes the form of tall thin columns aligned with the rotation axis that exhibit minimal variation in the $\zhat$ direction owing to the rotational constraint (Proudman-Taylor theorem) \citep{gastine2016scaling, long2020scaling, naskar2025force}. The stratification imposed by the F-layer represents a barrier to flow; convection columns crossing from the region outside the F-layer to the regions above and below the F-layer would have to significantly change their vertical length, which is not favoured at the low Rossby number conditions of the simulations. Therefore, convective flow structures tend to set up outside the cylinder that circumscribes the top of the stable region at its equator. These dominantly columnar flow structures appear at different latitudes at the top of the core, depending on the thickness and stratification strength of the F-layer. 

The expression of the stratified F-layer in the flow structure just below $\ro$ depends on $\rs$, $\dcidr$, and $\mathrm{Ra}_{T}$. Decreasing $\rs$, $\dcidr$, or increasing $\mathrm{Ra}_{T}$ reduces the thickness of the stable region and hence the offset between profiles of $u_r$ and $u_{\phi}$ in cases with and without an F-layer. Therefore, any signature of the stratified F-layer in observations of flow or magnetic field at the top of the core will be most prominent when the layer is thick and strongly stratified and will diminish as these properties are reduced.

\subsection{Dynamo Simulations}
\label{sec:results_mag}

\begin{figure}[h!]
\centering
\vspace{5mm}
\begin{overpic}[width=0.63\linewidth,trim={0cm 0cm 0cm 0cm},clip]{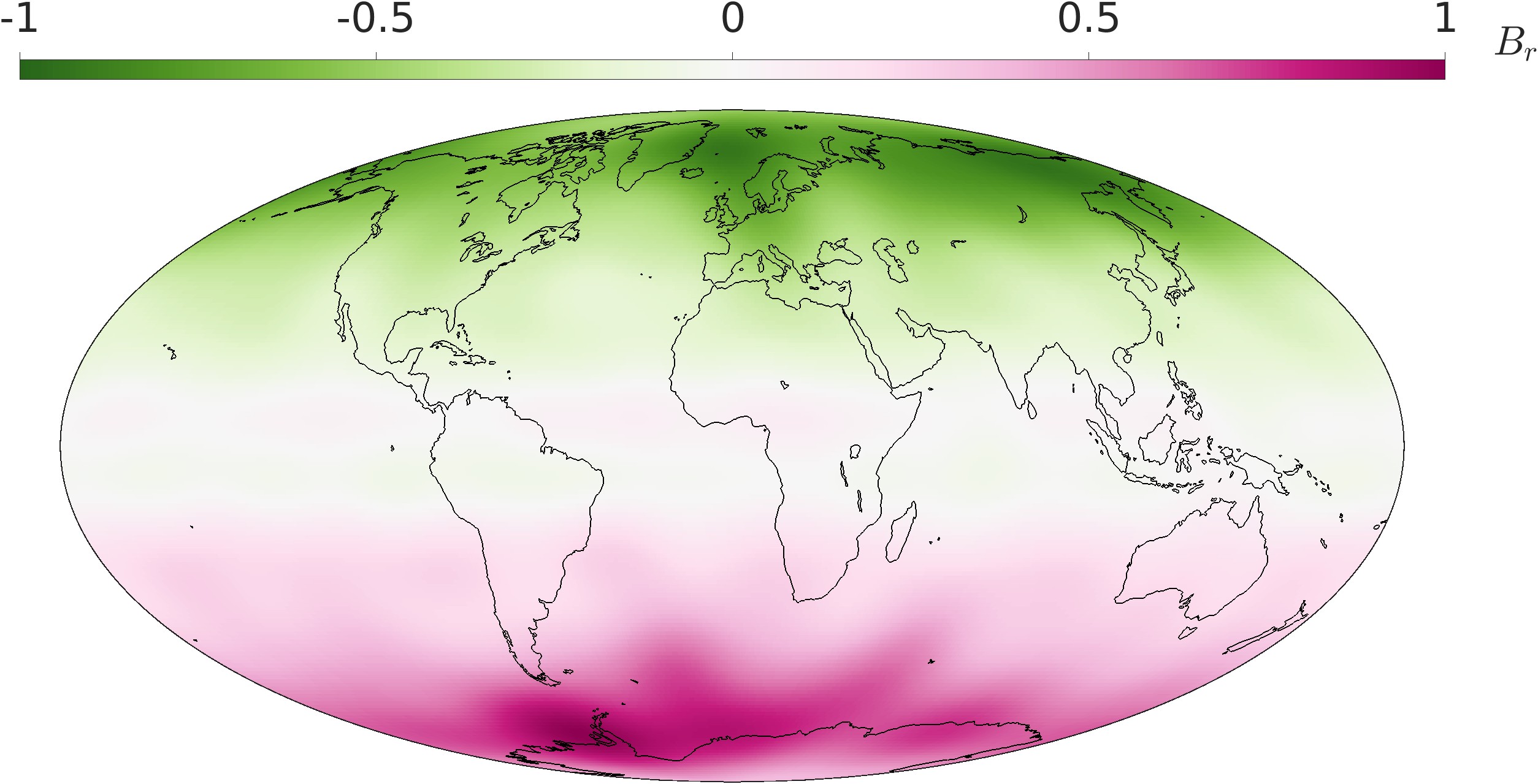}
\put(-2,55){(a) case DC02, $\dcidr=-0.4725$}
\end{overpic}
\begin{overpic}[width=0.25\linewidth,trim={0cm 0cm 0cm 0cm},clip]{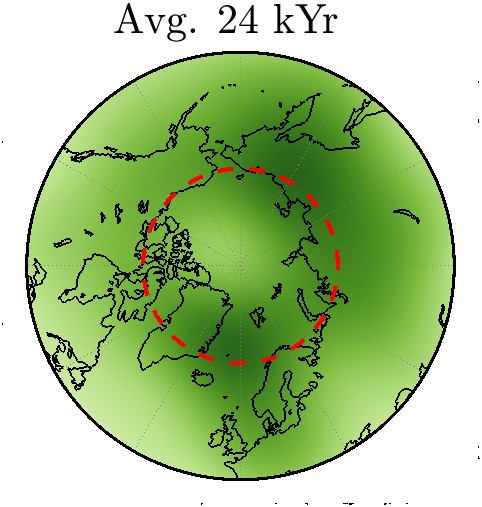}
\end{overpic}

\vspace{5mm}

\begin{overpic}[width=0.63\linewidth,trim={0cm 0cm 0cm 5cm},clip]{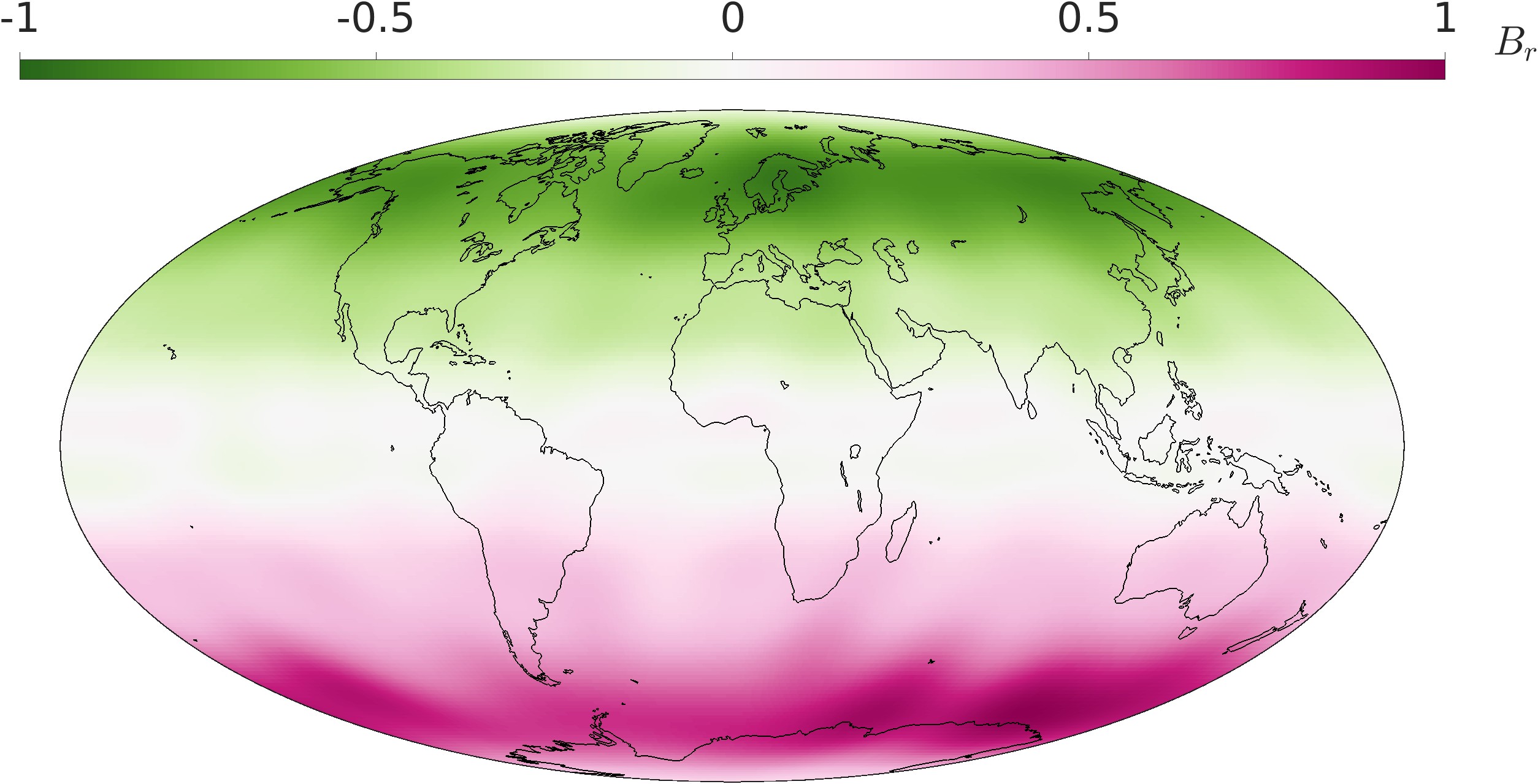} 
\put(-2,45){(b) case DC05, $\dcidr=20$}
\end{overpic}
\begin{overpic}[width=0.25\linewidth,trim={0cm 0cm 0cm 0cm},clip]{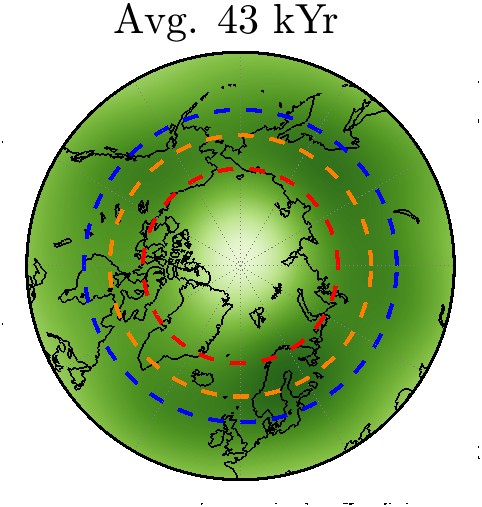}
\end{overpic}

\vspace{5mm}

\begin{overpic}[width=0.63\linewidth,trim={0cm 0cm 0cm 5cm},clip]{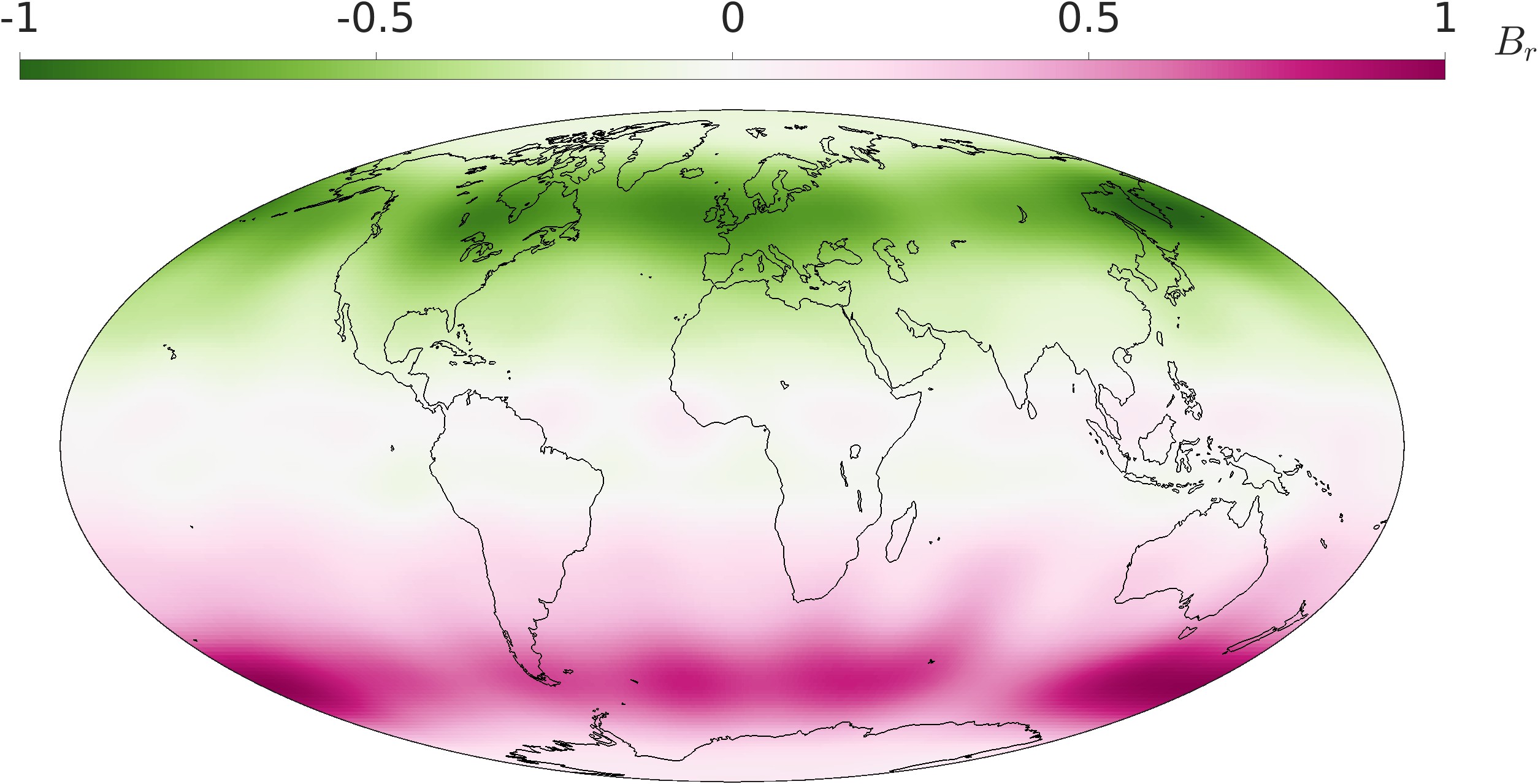}
\put(-2,45){(c) case DC08, $\dcidr=100$}
\end{overpic}
\begin{overpic}[width=0.25\linewidth,trim={0cm 0cm 0cm 0cm},clip]{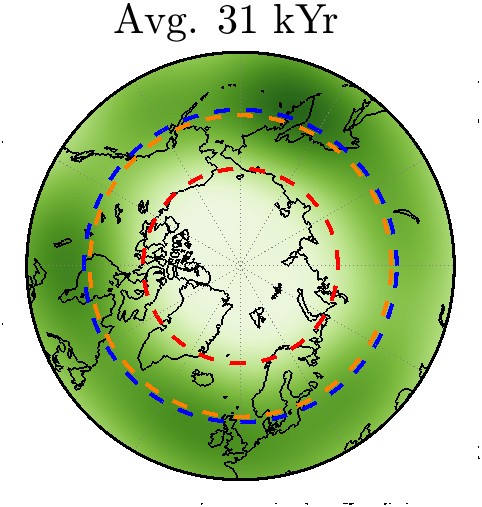}
\end{overpic}
\caption{Time-averaged radial magnetic field (normalised by its maximum) at $\ro$ truncated at $\ell_{\rm max}=14$. (a) No F-layer (case DC02), (b) F-layer (case DC05) with $\dcidr=20$ and (c) F-layer (case DC08) with $\dcidr=100$. DC05 and DC08 have $\mathcal{N}^2_\mathrm{i}=0.34$ and $\hf=389$~km, and $\mathcal{N}^2_\mathrm{i}=2.06$ and $\hf=612$~km, respectively. All other parameters are fixed: $\mathrm{Ra}_{T}=12 \times 10^7$, $r_{s}=0.83$. The red, orange, and blue circles in the right-hand panel mark the tangent cylinder, F-layer radius determined by the $\mathcal{N}^2 = 0$, and the radius $\rs$ in the reference state. Time has been rescaled to dimensional units using the magnetic diffusion time $D^2/\eta$ with $\eta = 1.6$~m$^{2}\,\textrm{s}^{-1}$. }
\label{fig:mag_bravg}
\end{figure}

Motivated by the results in section~\ref{sec:results_nonmag}, we begin in Fig.~\ref{fig:mag_bravg} by examining the time-averaged radial magnetic field on $\ro$ truncated at spherical harmonic degree $\ell_{\rm max}=12$, which is comparable to the resolution of historical geomagnetic field models. Three dynamo cases are denoted DC02, DC05, and DC08 with three different values of $\dcidr$, with all other parameters fixed (corresponding to the non-magnetic simulations C03, C28, and C34, respectively). All three cases are characterised by a non-reversing axial dipole-dominated time-average field. Case DC02, with no F-layer, shows a slight weakening of $B_r$ in the polar region compared to what would be produced from a purely axial dipole field, as has been found in many previous simulations \citep[e.g.][]{olson2002time, sreenivasan2006role}. As $\dcidr$ is increased, the size and relative weakening of this polar region increase, and a clear band of maximum $B_r$ emerges away from the pole in each hemisphere. 

\begin{figure}
    \centering
    \includegraphics[width=1\textwidth]{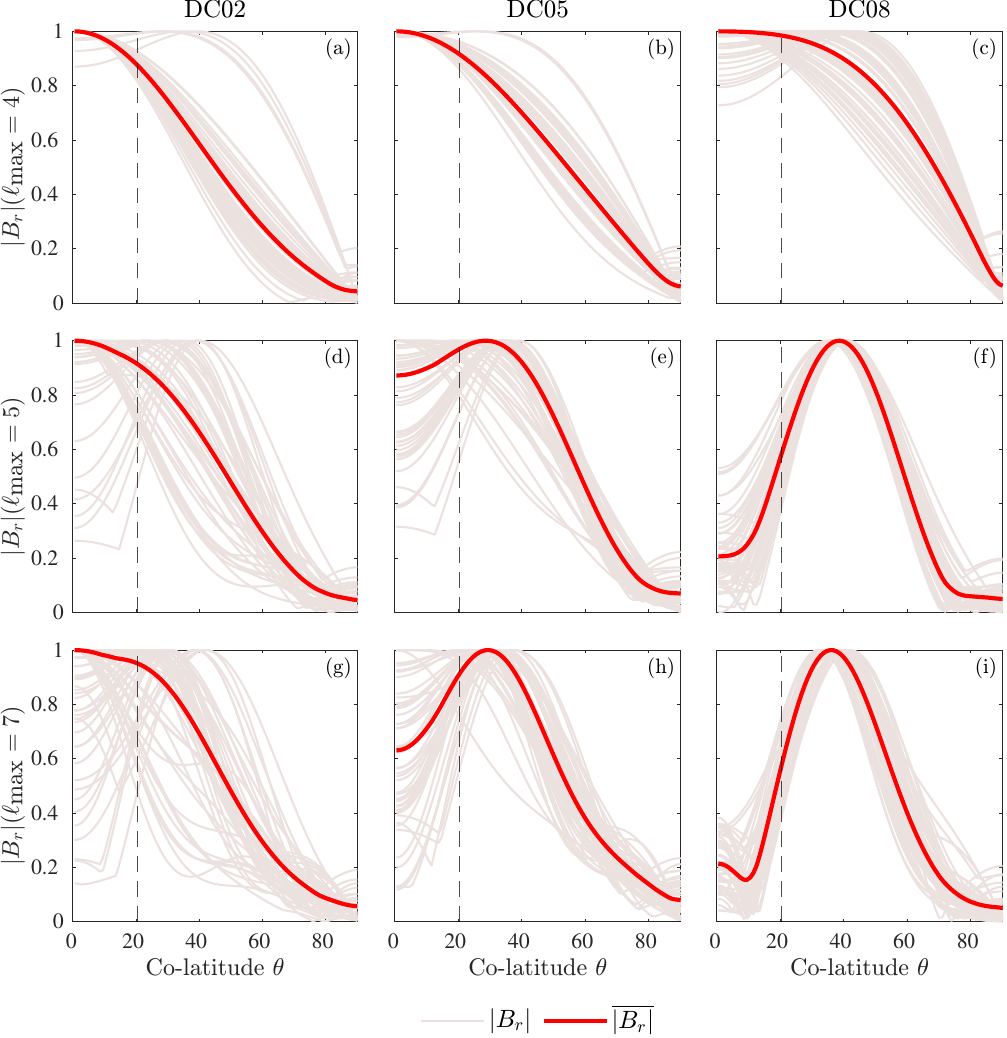}
    \caption{Latitudinal profiles of $|B_r|$ at $\ro$ averaged in time and azimuth and between northern and southern hemispheres for three simulations: DC02 (a,d,g); DC05 (b, e, h); and DC08 (c, f, i), with a truncation degree $\ell_{\rm max}$ of 4, 5, and 7 (rows). Profiles are normalised by their maximum. Reddish grey lines show 100 snapshots of the azimuthally averaged profile of $|B_r|$. The F-layer cases have $r_s=0.83$. The latitude of the tangent cylinder is highlighted by the dashed vertical line. }
    \label{fig:mag_brprof}
\end{figure}

In Fig.~\ref{fig:mag_brprof}, we investigate the weakening of $B_r$ in the polar regions in more detail by plotting time-averaged $|B_r|$ as a function of co-latitude for cases DC02 (no F-layer), DC05, and DC08. We show the profiles averaged between the North and South hemispheres, as the two hemispheres are indistinguishable in our simulations. We consider different spherical harmonic truncations from $\ell_{\rm max}=4$, to $\ell_{\rm max}=7$. At the surface, changing the truncation makes little difference; profiles of $\overline{|B_r|}$ for the three cases are essentially indistinguishable (see Supplementary Fig.~\ref{SI:mag_brprof_surf}). At the core mantle-boundary, with $\ell_{\rm max}=4$, as $\dcidr$ increases, there is an increase in $|B_r|$ at all latitudes.  and also an increase in the variability of $|B_r|$ (shown by instantaneous profiles in pink). Increasing $\ell_{\rm max}$ to 5 reveals only marginal changes to the profile for case DC02, but the emergence of polar minima for cases DC05 and DC08. The profile of case DC08 is particularly striking, with a strong peak in $\overline{|B_r|}$ around $40^{\circ}$ co-latitude and a polar minimum of comparable intensity to the equatorial field strength. Upon increasing $\ell_{\rm max}$ to 7, there is a barely discernible polar minimum in case DC02, while the polar minimum in case DC05 is more clearly defined. There is little change in the profile for case DC08 between $\ell_{\rm max}=5$ and $\ell_{\rm max}=7$. In DC02 (no F-layer case), a peak in $|B_r|$ and a polar minimum can be observed in snapshots but not in the overall time-average. Indeed, the peak of $\overline{|B_r|}$ is only distinguishable with $\ell_{\rm max}>9$ (see Supplementary Fig.~\ref{SI:mag_brprof14}), while $\ell_{\rm max}>5$ is enough for the F-layer case. While the mean position of the maximum of $\overline{|B_r|}$ and the polar minimum converge when increasing $\ell_{\rm max}$, instantaneous profiles of $|B_r|$ are more variable, especially for no F-layer simulations compared to F-layer ones.

\begin{figure}
    \centering
    \includegraphics[width=1\textwidth]{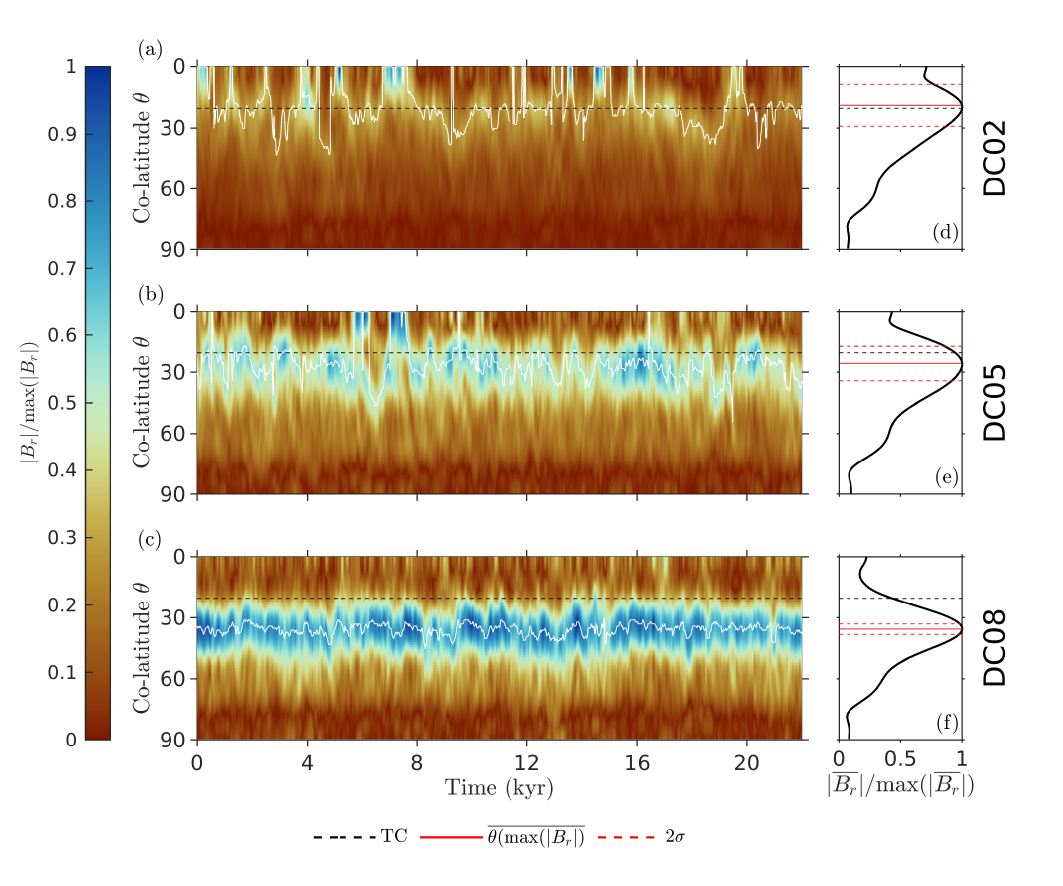}
    \caption{(a, b, c) Spatio-temporal diagram of $|B_r|$ at $\ro$ for three different dynamo simulations with $\mathrm{Ra}_{T}=12 \times 10^7$ and $(\dcidr=-0.4725;\, r_s=0)$, $(\dcidr=20;\, r_s=0.83)$, $(\dcidr=100;\, r_s=0.83)$ respectively. $B_r$ is obtained from the simulations with a truncation at $\ell_{\rm max}=14$ and averaged between the North and South hemispheres. The white lines show the position of the peak of $|B_r|$. (d, e, f) Time-average of the profile of $|B_r|$.  The dashed horizontal line shows the co-latitude of the tangent cylinder. Red (solid and dashed) lines are the mean of the maximum of $|B_r|$ (from a-d) and its standard deviation ($2\sigma$), respectively. The position of the maximum of $|B_r|$ in (d, e, f) is also shown in Fig.~\ref{fig:max_br}.}
    \label{fig:mag_spatiotemp_br}
\end{figure}

To better understand the temporal variability of the magnetic field, Fig.~\ref{fig:mag_spatiotemp_br} shows time-latitude plots of $|B_r|$ at $\ro$ for the three cases of Fig.~\ref{fig:mag_bravg} along with the latitudinal profiles of $|B_r|$ averaged in time and azimuth. In case DC02, the maxima in $|B_r|$ generally appear near the tangent cylinder, though there are significant deviations, e.g., around $t=7.3$~kyr where the maximum is near the pole, and around $t=9.5$~kyr where it moves to $\sim40^{\circ}$ latitude. In this case, the mean location of $\max{|B_r|}$ is $\theta = 19^{\circ}$ and the $\pm 2\sigma$ bounds lie in the range $\theta = 8.8^{\circ}-29.2^{\circ}$. In case DC05, peaks in $|B_r|$ occur at lower latitudes and in a narrower latitude band, though occasionally they are seen near the poles (e.g., around $t=6$~kyr or $t=7.3$~kyr). In case DC08, the peaks in $|B_r|$ are always confined to the band $\theta = 33.2^{\circ}-38.3^{\circ}$ ($2\sigma$) and are never seen near the poles or at low latitudes. 
This raises the interesting possibility that bounds on the combined F-layer thickness and stratification strength can be obtained from the 400-yr historical geomagnetic field record, which has sufficient spatial resolution to resolve the predicted polar minimum (Fig.~\ref{fig:mag_brprof}). 

It is important to recognise that the presence of a polar minimum (reciprocally a maximum) in $|B_r|$ does not, on its own, represent a geomagnetic signature of the stratified F-layer, as we can observe it in no F-layer simulations (e.g. DC02). Previous simulations without an F-layer have shown that polar minima depend on the flow inside the tangent cylinder and that minima tend to decrease as the Ekman number decreases \citep[e.g.][]{cao2018geomagnetic,lezin2023mantle}.  However, these polar minima are generally confined within the tangent cylinder (see DC02 or simulations in \citet{olson2002time}, \citet{sreenivasan2006role}, or \citet{lezin2023mantle}). The crucial effect of a strongly stratified and thick F-layer in our simulations is to produce a strong polar minimum, with strength similar to that of the equatorial field, and to move the peak in $|B_r|$ to lower latitudes with little temporal variability. It is these combined effects that can be distinguished with geomagnetic observations.

\subsection{Comparison against observations}

Here, we will compare the magnetic field produced by our simulations to three geomagnetic fields models: GGF100k, which spans the last 100~kyrs with a spatial resolution of $\ell_{\rm max} \approx 4$ and temporal knot spacing of 200 yrs \citep{panovska2019one}; CALS10k.2, which spans the last 10~kyrs with a spatial resolution of $\ell_{\rm max} \approx 4$ and temporal knot spacing of 40 yrs \citep{cals10k2}; and the gufm1 model spanning the historical period with spatial resolution of $\ell_{\rm max} \approx 14$ and temporal knot spacing of 10 yrs \citep{jackson2000four}. First, we compare the polar minimum and the position of the maximum of $|B_r|$ to the gufm1 model, as higher resolution is required to distinguish a peak in the profile of $B_r$. To investigate longer timescales where the signature of the F-layer might more readily emerge on averaging, we must contend with lower spatial resolution. We therefore also compare the simulations and field models in terms of ratios of low-degree time-averaged Gauss coefficients.

\begin{figure}
    \centering
    \includegraphics[width=1\textwidth]{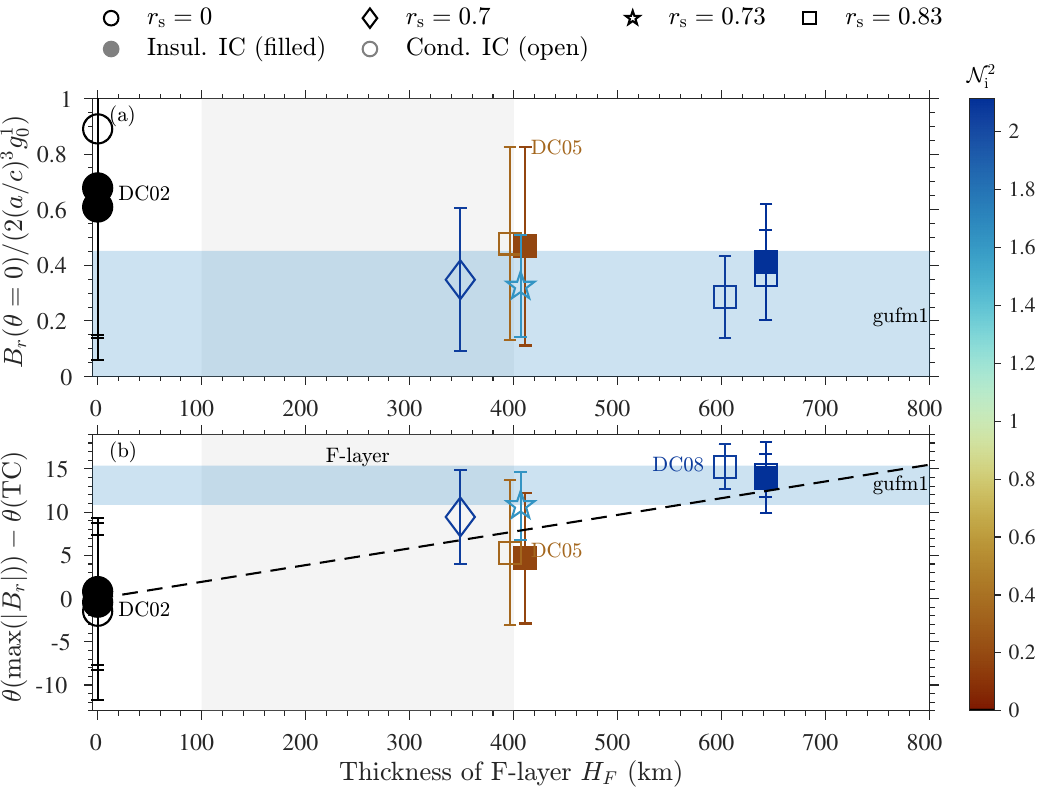}
\caption{(a) Polar minimum intensity (normalised by dipole component of $g_0^1$ at the pole, with $a$ and $c$, the radius of Earth and Earth's core) as a function of the thickness of the stable layer (F-layer). (b) Co-latitudinal distance between the maximum of $|B_r|$ and the tangent cylinder of the inner core as a function of the thickness of the stable layer (F-layer). Filled and empty symbols denote dynamo simulations with an insulating or conductive inner core, respectively. Right triangles, diamonds, pentagrams, and squares denote $r_s= 0.70,\,0.73,\, 0.83,\, 1.03$, respectively. Black circles denote no F-layer cases. Colours show the different values of the strength of the stratification $\mathcal{N}^2_\mathrm{i}$. The black dashed line denotes the position of the tangent cylinder of the F-layer relative to the inner core tangent cylinder. Vertical grey area denotes the plausible thickness of the F-layer seen by seismic observations \citep{souriau1991velocity,kennett1995constraints,ohtaki2018seismological, adam2018observation, ohtaki2025seismological}. Horizontal blue areas denote $2\sigma$ of polar minima or the position of the maximum of $|Br|$ for gufm1 ($\ell_{\rm max}=14$, \cite{jackson2000four}). Note that $\theta(\textrm{max}(|B_r|)) -\theta(\textrm{TC})=0$ and $\sim -21^\circ$ correspond to the tangent cylinder and to the pole, respectively. The size of error bars represents $2\sigma$.}
\label{fig:max_br}
\end{figure}

Fig.~\ref{fig:max_br} shows the intensity of the polar minimum normalised by the polar value of the dipole component and the co-latitudinal distance of the peak in $|B_r(\ro)|$ from the TC as a function of $\hf$ for all dynamo simulations truncated at $\ell_{\rm max} =14$. As seen in Fig.~\ref{fig:mag_brprof}, the polar minimum decreases while increasing the thickness or the strength of the F-layer stratification. While the time-averaged value of our no F-layer simulation lies just outside the $2\sigma$ of gufm1, the temporal mean for all of our F-layer simulations is consistent with the polar minimum observed in gufm1. As is the case for the downwelling position in non-magnetic simulations, increasing $\dcidr$ and $\rs$ moves the peak of $|B_r|$ to lower latitude as $\hf$ increases. Again, there is a trade-off between the parameters that control the F-layer; for example, simulations with $r_s=0.83$ and weak stratifications (low $\mathcal{N}_i^2$) have peaks of $|B_r|$ at similar latitudes to the simulation with $r_s=0.73$ and a stronger stratification. 
We ran simulations with conducting or insulating inner core with the same parameters, appearing as empty and filled markers, respectively. This modification to the simulation setup does not significantly alter the influence of the F-layer on $|B_r(\ro)|$.

We also show in Fig. \ref{fig:max_br}(b) the mean of the position of the peak of $|B_r|$ (and its $2\sigma$ of standard deviation) in the Northern hemisphere of gufm1 and the range of F-layer thicknesses deduced from seismic observations. All simulations that include an F-layer overlap gufm1 within the $2\sigma$ range. This is notably true for simulations with F-layer thickness within the thickness range expected from seismology. Although it is not the case for simulations without an F-layer, some periods, spanning between $\sim 10 \textrm{ to } 700$~yrs, do have positions of the peak of $|B_r|$ consistent with gufm1. Simulations with strong stratifications ($\mathcal{N}^2>1$) and F-layer thicknesses larger than expected from seismology fit gufm1 the best, although with low time variability, suggesting that a very thick F-layer cannot be excluded based on the last 400 years of magnetic observations. It also suggests that simulations with similar behaviour to ours will not simultaneously match the time-averaged $|B_r(\ro)|$ profile from gufm1 and the seismically-inferred F-layer thickness. This means that either gufm1's $|B_r(\ro)|$ profile is not indicative of the time-averaged geomagnetic behaviour or that a more refined tuning of simulation parameters is required to match the observations. 

\begin{figure}
    \centering
    \includegraphics[width=1\textwidth]{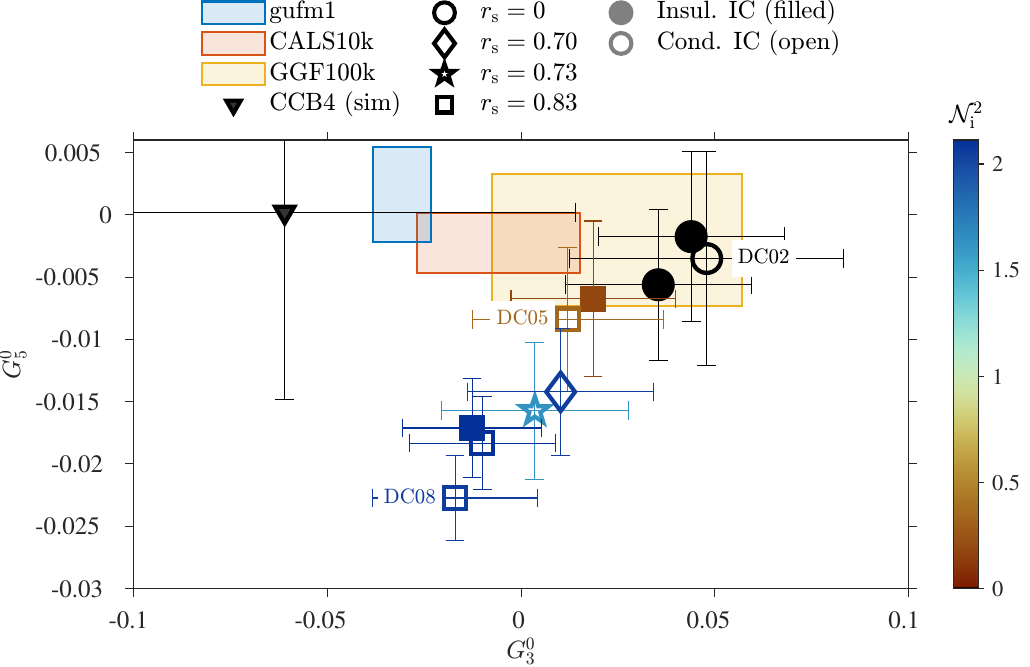}
    \caption{Gauss coefficient ratios $G_3^0$ and $G_5^0$ of the no F-layer and F-layer simulations compared with field models. Symbols and colours are similar to those in Fig.~\ref{fig:max_br}. 
    Black downward triangle denotes low inertia reversing dynamo simulation CCB4 from \cite{jt2025}. Blue, red, and yellow boxes show gufm1 \citep{jackson2000four}, CALS10k.2 \citep{cals10k2}, and GGF100k \citep{panovska2019one} field models, respectively. The size of the boxes and error bars represents $2\sigma$.}
    \label{fig:gauss_ratios}
\end{figure}

The long-term evolution of the profile of $|B_r|$ shows significant differences in terms of variability between no F-layer and F-layer cases. The anomalous structure in $\overline{|B_r|}$ at $\ro$ induced by the stratified F-layer (Fig.~\ref{fig:mag_brprof}) is most clearly identified in equatorially antisymmetric terms in the spherical harmonic expansion of $B_r$. Fig.~\ref{fig:gauss_ratios} shows the time-averaged ratio $G_5^0 = g_5^0/g_1^0$ plotted against $G_3^0 = g_3^0/g_1^0$ for all of our dynamo simulations. The value and variability of $G_3^0$ and $G_5^0$ decrease with increasing $\dcidr$. While different combinations of low-degree zonal coefficients can produce similar profiles of $|B_r|$, the reduction in $G_3^0$ is consistent with the increase in $|B_r|$ at mid-latitudes seen in Fig.~\ref{fig:mag_brprof} (top row), and the enhanced negative $G_5^0$ term acts to reduce the field strength near the poles and enhance $B_r$ in the mid-latitudes (specifically, in the range $30^{\circ}-60^{\circ}$; Fig.~\ref{fig:mag_brprof} middle row). The trend arising from Fig.~\ref{fig:gauss_ratios} is that both $G_3^0$ and $G_5^0$ decrease and become negative as the F-layer becomes thicker and more strongly stratified. 

Fig.~\ref{fig:gauss_ratios} also shows time-averaged $G_3^0$ and $G_5^0$ for geomagnetic field models gufm1, CALS10k.2 and GGF100k. Taken at face value, simulations with no F-layer, or a weakly stratified F-layer, are closest to matching the observations (within $2\sigma$). Increasing $\rs$ and $\dcidr$ reduces both ratios to negative values, which improves the agreement with $G_3^0$ from gufm1 and CALS10k.2 but reduces the agreement with the $G_5^0$ from these models. None of the simulations matches gufm1 or CALS10k.2 in time-average, and most do not match them at individual instants in time. The main reason is that our homogeneous simulations show a strongly positive $G_3^0$. This positive time-averaged $G_3^0$ is often but not always found in homogeneous dynamo simulations, which is illustrated by the inclusion in Fig.~\ref{fig:gauss_ratios} of a homogeneous simulation from \cite{jt2025}. Therefore, the diagnostic of the F-layer is not the absolute value of $G_3^0$ (or $G_5^0$) but rather the trend of decreasing $G_3^0$ and $G_5^0$ with increasing $\rs$ and $\dcidr$. 

In addition, we have also computed the morphological compliance criteria \citep{christensen2010conditions} and the paleomagnetic Q$_{\rm PM}$ criteria  \citep{sprain2019assessment} for each of our simulations (Table 1 in Supplementary dataset \citep{hfnmcrmd2026}). These criteria are designed to quantitatively compare the field morphology and paleomagnetic properties of simulated and observed fields, respectively. In the time-average, our simulations are moderately compliant with the modern geomagnetic field morphology ($\chi^2$ values of \citet{christensen2010conditions} in the range 2.90--6.90), while individual snapshots in each simulation are in excellent agreement with the modern field ($\chi^2 < 2$). We find no systematic variation of the morphological properties identified by \citet{christensen2010conditions} with $\rs$ or $\dcidr$. In terms of Q$_{\rm PM}$, the simulations do not show strong agreement with paleomagnetic observations from the last 10 Myrs, in part because they are not long enough and do not produce excursions or reversals. Nevertheless, we do observe a systematic reduction in the median and variance of virtual dipole moment (VDM) variability with increasing $\rs$ and $\dcidr$. We compare VDM variability to GGF100k, which is of similar length to our simulations and has  $V\% = IQR(VDM)/med(VDM)$ = 0.306. The simulations that most closely reproduce this value tend to have no or thinner and weakly stratified F-layers. Variations in the other components of the Q$_{\rm PM}$ criteria show no correlation with the degree of stratification ($\rs$ and $\dcidr$).

\section{Discussion and Conclusions}
\label{sec:discussion}

We have studied numerical simulations of rotating spherical shell thermochemically-driven convection and dynamo action with a chemically stratified region at the base of the liquid core that provides a simple representation of the slurry F-layer model proposed by \citet{wong2021regime} and \citet{wilczynski2025two}. The simulated F-layer is governed by its imposed initial thickness $\rs$, stratification strength at the lower boundary $\dcidr$, and the amplitude of its destabilising thermal buoyancy $\mathrm{Ra}_{T}$. We find that increasing $\rs$, $\dcidr$, and $\mathrm{Ra}_{C}/\mathrm{Ra}_{T}$ increases the stratification strength and thickness of the F-layer. With sufficiently strong stratification, the F-layer acts as a barrier to quasi-geostrophic flow and the dominant convective structures are displaced to larger (cylindrical) radii, an effect that is visible near the top of the core. This effect increases with increasing $\rs$ and $\dcidr$ as the F-layer is thickened and becomes more stratified, and when increasing $\mathrm{Ra}_{C}/\mathrm{Ra}_{T}$ as the relative amplitude of the destabilising thermal convection is reduced. In the magnetic field, this translates into a deepening of polar minima at the core surface to a level that can be comparable to the equatorial field. The peak in latitudinal profiles in the radial magnetic field at the core surface is also pushed towards lower latitudes with reduced temporal variability. Given the effect of the F-layer on the CMB magnetic field, geomagnetic observations may provide independent constraints on the F-layer. 

Our simulations employ a number of simplifications to represent the F-layer and the dynamics of the convecting core. \citet{wilczynski2025two}'s slurry model of the F-layer is fully compressible and requires the presence of both solid and liquid phases to produce stable stratification. Our Boussinesq single-phase simulations can therefore only approximate the radial profiles of temperature, composition, and liquid velocity predicted by the \citet{wilczynski2025two} model. The effect of these simplifications is hard to assess because the slurry equations solved by \citet{wilczynski2025two} are also subject to simplifying assumptions; in particular, solutions are 1D, steady, and ignore rotation and magnetic fields. At geophysically relevant conditions, the slurry model produces a temperature profile that is linear (following the liquidus) and destabilising, a compositional profile that is close to linear and stabilising, and constant vertical liquid velocity through the bulk of the layer. Our imposed thermal and chemical profiles are therefore similar to \citet{wilczynski2025two}'s solutions. The velocity must increase to the amplitude of the bulk across a region near the top of the F-layer that is not modelled by \citet{wilczynski2025two}; however, for our simulations, this region is rather broad (owing to the cost of its numerical representation) and, so we do not see convincing evidence of a uniform velocity profile in the F-layer liquid. The suppression of fluid motion in the F-layer compared to the bulk core is also weaker in our simulations than expected from the slurry model, although we do achieve strongly stratified layers with $\mathcal{N}^2\sim 1$ and radial velocity reductions of $2-4$ orders of magnitude. Overall, we believe that the simplified F-layer representation adopted in this study represents a sensible compromise between accuracy and practicality. Indeed, it could be argued that these simplifications are desirable since they embody the essential ingredients that are necessary to explain seismic observations of the F-layer --- thermally destabilising gradients and a stabilising compositional component \citep{gubbins2008thermochemical} --- without being too heavily tied to the slurry model. 

The representation of core dynamics in our simulations is compromised due to computational limitations in representing the range of spatio-temporal scales that arise in Earth's core. Our choice of Ekman number $\Ek = 2\times 10^{-5}$, compared to Earth's core value of $10^{-15}$ \citep{davies2015constraints}, is greater than the state-of-the-art direct numerical simulations \citep{schaeffer2017turbulent, lin2025invariance}, but is low enough to access quasi-geostrophic dynamics and high enough to allow variation of the F-layer properties, which is the focus of this work. The choices $\mathrm{Pr}_{T}=1$ and $\mathrm{Pr}_{C}=10$ are, respectively, higher and lower than values for Earth's core \citep{davies2015constraints}, but are comparable to recent studies of top-heavy double diffusion in rotating spherical shells \citep{tassin2021geomagnetic}. The range of Rayleigh numbers is then set to obtain dipole-dominated solutions, which in our simulations do not reverse polarity. We emphasise again that $\mathrm{Ra}_{T}\ll \mathrm{Ra}_{C}$ both for Earth's bulk core and the F-layer, and so we focus on simulations where this inequality is best satisfied when interpreting the simulation results in the context of the Earth. 

The fact that we have considered only one value of $\Ek$, $\mathrm{Pr}_{T}$, $\mathrm{Pr}_{C}$, and $\mathrm{Ra}_{C}$ limits our capacity to quantitatively extrapolate our simulation results to Earth's core conditions, and we do not attempt this exercise. Instead, we note that dipole-dominated dynamos with homogeneous boundary conditions invariably produce time-averaged outer boundary fields that are dominantly axisymmetric and with the latitudinal maximum in $B_r$ that is within the tangent cylinder \citep{olson2002time, davies2011buoyancy, driscoll2016simulating,landeau2017signature, olson2017dynamo}. This feature of the simulations is thought to reflect the theoretically expected distinction in dynamics inside and outside the TC in Earth's core. We therefore believe that the effects of the F-layer revealed in the time-averaged radial outer boundary field could persist to more geophysically realistic conditions. 

Our simulations suggest that peaks in $|B_r|$ at the core surface systematically move to lower latitudes and exhibit reduced temporal fluctuations as F-layer thickness and stratification strength are increased. For the most strongly stratified cases, peaks in $|B_r|$ are never found within the tangent cylinder across the whole timespan of our simulations, which is inconsistent with behaviour seen in models of the historical magnetic field. This suggests that relatively short timescale global geomagnetic field models, such as gufm1, can be used to constrain some properties of the F-layer. This is desirable since these models have sufficient spatial resolution to constrain the polar minima of the core-mantle boundary field, which is not possible with the more limited resolution of Holocene and longer timescale global field models. However, from our simulations, this constraint on plausible temporal variability only rules out F-layers of around 600~km, which is already thicker than layers reported from seismic studies. Therefore, while providing novel insight, it is only a loose constraint on F-layer properties. 

We find a systematic decrease in the Gauss coefficient ratios $g_3^0/g_1^0$ and $g_5^0/g_1^0$ of the surface field. In our homogeneous simulations, $g_3^0/g_1^0$ is positive and larger than values obtained for the historical and Holocene geomagnetic field, while $g_5^0/g_1^0$ is negative and comparable to values from these models. \citet{yan2018sensitivity} showed that the addition of a stratified layer below the outer boundary reduces $g_3^0/g_1^0$ and that matching the observed values could be achieved for different combinations of layer thickness and stratification strength.  However, the existence of a stable region at the top of Earth's core is still debated \citep{Gubbins07a, Alexandrakis10, Lesur15, vanTent20, davies2023dynamics, aubert2025core, aubert2025rapid}. Our results suggest that the presence of a stratified F-layer can also reduce $g_3^0/g_1^0$ to within the range inferred from geomagnetic field models spanning a range of timescales. However, we caution that the ratio $g_3^0/g_1^0$ is also strongly dependent on the parameters (e.g., Rayleigh number, Prandtl numbers) that control the homogeneous dynamo. Therefore, explaining the $g_3^0/g_1^0$ ratio of Earth's modern field in terms of anomalous layering in the core requires future work that examines a much broader range of simulations than have been conducted here. 

The main insight drawn from our simulations is that geomagnetic observations can potentially constrain the thickness and stratification strength of the F-layer. The signature of the F-layer in $B_r$ at the core surface is clearest in the time-averaged field, which emerges in a few tens of kyrs in our simulations, a timespan that is covered by the GGF100k global time-dependent reconstruction of the last 100 kyrs. This model does not currently have sufficient resolution to capture the strong polar minima in $B_r$ that emerge for a spherical harmonic truncation of at least degree 5, but future improvements may allow degree 5 structure to be resolved, which may provide a strong constraint on the properties of the F-layer according to our simulations. Future work should also investigate the role of the F-layer in a wider range of homogeneous simulation setups to better constrain its role in determining absolute time-averaged values of low-degree Gauss coefficients and virtual dipole moment variability. The impact of heterogeneous heat flux at the core–mantle boundary on the structure of the F-layer should also be evaluated in light of the laterally varying F-layer properties inferred from seismology \citep{ohtaki2018seismological,ohtaki2025seismological}, since our results indicate that lateral variations remain relatively small when compositional convection is strongly dominant, as is expected in Earth’s core. Our findings do not support an F-layer that is both thick ($>600$~km) and strongly stratified (normalised Brunt–Väisälä frequency $>1$). Instead, based on our simulations and comparison with available observations, we favour a weakly stratified ($\mathcal{N}^2<1$) F-layer with a thickness of less than $400$~km. These constraints offer additional insight into the nature of the F-layer and serve as a useful complement to seismic observations.

\section*{Acknowledgements}

We gratefully acknowledge support from Natural Environment Research Council grants NE/Y003500/1 (supporting CD and SM), NE/V010867/1 (supporting CD, AC and HR), NE/W005247/1 (supporting JM and SN), and NE/X014142/1 (supporting TF). We are also grateful to the Science and Technologies Facilities Council for grant ST/Y001206/1, which supports LH, JM, and CD. Calculations were performed on the UK National supercomputing service ARCHER2 and on the University of Leeds service Aire.

\clearpage
\newpage
\appendix
\section{Supplementary material}
\setcounter{figure}{0}
\renewcommand{\thefigure}{S\arabic{figure}}
\setcounter{equation}{0}
\renewcommand{\theequation}{S\arabic{equation}}
A dataset with three tables is provided \citep{hfnmcrmd2026}. First, Table 1 provides the input parameters ($\Ek$, $\mathrm{Pm}$, $\mathrm{Pr}_C$, $\mathrm{Pr}_T$, $\mathrm{Ra}_{T}$, $\mathrm{Ra}_{c}$, $\dcidr$, $\rs$, $\chi$) defined in the main text and all main diagnostic parameters characterizing the thermochemical and dynamo simulations ($\mathrm{{Rm}}$, $\mathrm{{Ro}}$, $\mathrm{{Ro}_l}$, $\Lambda$, $M=\magEnergy/\kinEnergy$, $D_{O}$, $f_\text{ohm}$, $f_\mathrm{dip}$) defined in the following section.

We define the following dimensionless output diagnostics to analyse the simulations. The kinetic and magnetic energies are defined as 
\begin{linenomath*}
\begin{equation}
    \kinEnergy = \frac{1}{2} \int_V \vel^2 \dd V , \quad \magEnergy = \frac{\Pm}{\Ek} \int_V \magf^2 \dd V ,
\end{equation}
\end{linenomath*}
where $V$ is the shell volume. The magnetic Reynolds number $\Rm$, the ratio of advection to diffusion of magnetic field, is $\Rm = \sqrt{2 \kinEnergy/V}$; the Elsasser number $\elsasser$, a measure of the ratio of Lorentz and Coriolis forces, is $\elsasser = (\magEnergy \Ek)/(\Pm V)$. 
From these definitions, we obtain the Rossby number $\Ro = \Rm/\Pm \Ek$ and local Rossby number $\mathrm{Ro}_\ell=\mathrm{Ro} \ell_u /\pi$ \citep{oa2014}.

\citep{oa2014}.
The ohmic dissipation fraction is calculated as 
\begin{linenomath*}
\begin{equation}
    f_\text{ohm}=\frac{D_{O}}{(D_{O} + D_{\nu})} ,
\end{equation}
\end{linenomath*}
where
\begin{equation}    
    D_{\rm \nu} = \Pm \int \left( \nabla \times \vel \right)^2 \mathrm{d} V , \quad
    D_O = \frac{2 Pm}{E} \int \left( \nabla \times \magf \right)^2 \mathrm{d} V .
\end{equation}
The dipolarity of the magnetic field is calculated using the standard formula \citep{christensen2006scaling} 
\begin{linenomath*}
\begin{equation}
 f_\mathrm{dip} = \left(
    \frac{\int_{S} \magf_{\ell=1}(r=\ro)\cdot\magf_{\ell=1}(r=\ro)\dd A}
    {\int_{S} \magf_{\ell \leq 12}(r=\ro)\cdot\magf_{\ell\leq 12}(r=\ro)\dd A} 
    \right)^{1/2},
\end{equation}
\end{linenomath*}
where $S$ is the surface of the CMB and $\ell$ is the spherical harmonic degree.\\

In Table 2, we provide data from the analysis of all simulations. The thickness of the stable layer is determined by the radius where the Brunt-Väisälä frequency is zero, using the temporal average of the radial profile of the Brunt-Väisälä frequency or the radial profile of the temporal average of the Brunt-Väisälä frequency. The squared dimensionless Brunt-Väisälä frequency at the inner core boundary is $\mathcal{N}^2_\mathrm{i}$ and the radial average of the squared dimensionless Brunt-Väisälä frequency between $\ri$ and $r(\mathcal{N}^2=0)$ is $<\mathcal{N}^2>$. This table also gives the time-averaged colatitude of the minimum of $u_r$, and the maximum of $|B_r|$ (and its standard deviation); the time-averaged of intensity of the polar minimum of $|B_r|$ and its standard deviation\\

In Table 3, we provide the time average of the Gauss coefficient $g_0^1$ and the ratio $G_0^X=g_0^X/g_0^1$ with $X$ up to degree 9. We additionally provide all Gauss coefficients $g_m^l$ and $h_m^l$ up to degree 3. The standard deviation $\sigma$ is also provided. Same values is provided for the three geomagnetic field models used in the paper gufm1 \citep{jackson2000four}, CALS10k.2 \citep{cals10k2}, GGF100k \citep{panovska2019one}.

\section{Force balance}
We evaluate the dynamical balance in our simulations from the volume-averaged scale-dependent forces in the momentum equation (equation \ref{eq:momentum}), following \citet{naskar2025force}. All the forces are integrated over the bulk fluid, which excludes regions of radial thickness $10$ times the Ekman layer depth adjacent to the upper and lower boundaries (defined based on the linear intersection method). In \citet{naskar2025force}, the forces and curled forces are partitioned into mean (i.e. azimuthal average) and corresponding fluctuating parts, and we only consider the fluctuating part of the forces here (indicated by a prime). The Inertia ($I$), Viscous ($V$), Coriolis ($C$), , Pressure gradient ($P$), thermal buoyancy (i.e., Archimedean force, $A_{T}$), compositional buoyancy ($A_{\xi}$) and the Lorentz force ($M$) magnitudes are defined as,
\begin{equation}
\begin{split}
     I =|(\boldsymbol{v^{'}}\cdot\boldsymbol{\nabla})\boldsymbol{v^{'}}| , ~~~
     C =  \frac{2\Pm}{\Ek}|\left(\boldsymbol{\hat{z}}\times\boldsymbol{v^{'}}\right)|, ~~~ 
     P = |\nabla\left(\frac{\mathrm{Pm}}{\Ek}\,\tilde{P} + \frac{1}{2}\lvert \vel \rvert^{2}\right)|, ~~~
     M =   \frac{2\Pm}{\Ek}|\left(\boldsymbol{\nabla}\times\boldsymbol{B^{'}}\right)\times\boldsymbol{B^{'}}|,\\
     A_{T} = \Pm^2|\left(\frac{Ra_{T}}{Pr_T}T\right)\radiusvec| , ~~~
     A_{\xi} = \Pm^2|\left(\frac{Ra_{C}}{Pr_C}C\right)\radiusvec| , ~~~  
     V = \Pm|\nabla^{2} \vel^{'}| , ~~~
\end{split}
\end{equation}

where the residual of the Coriolis and pressure forces is termed as ageostrophic Coriolis $C'_{ag}$, and $|.|$ represents the r.m.s magnitude of forces. 

\begin{figure}
    \centering   
    \begin{overpic}[width=0.4\textwidth]
    {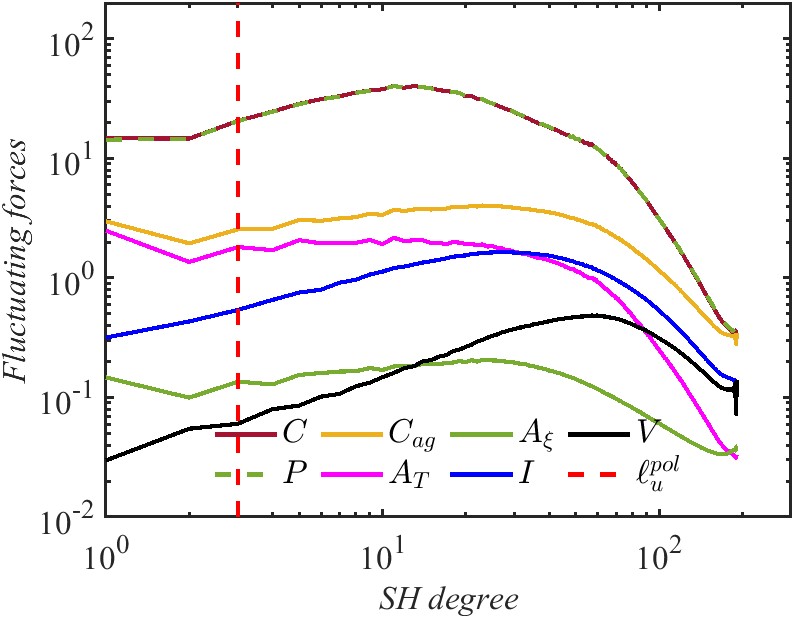}
    \put(0,80){$(a)\ \textrm{case}\ \textrm{C03}$}
    \end{overpic}\\
    \vspace{2mm}
    \begin{overpic}[width=0.4\textwidth]
    {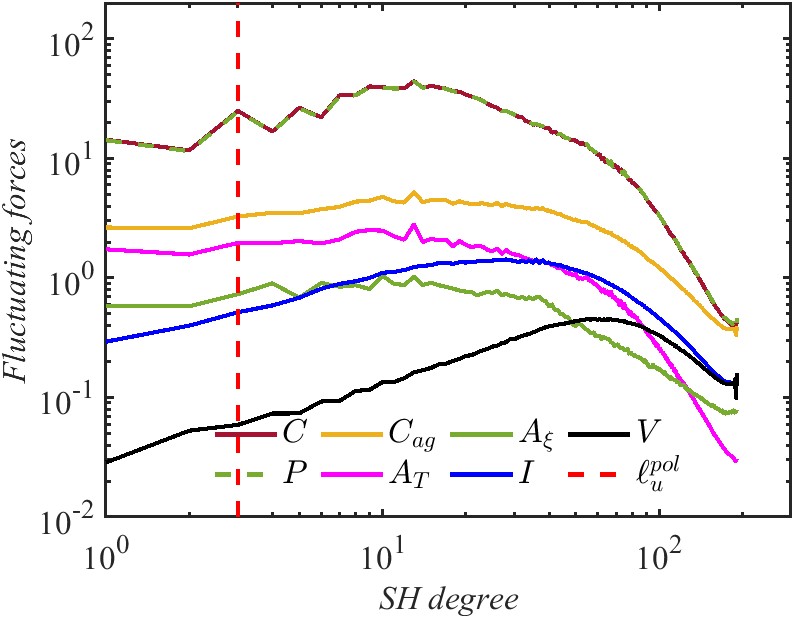}
    \put(0,80){$(b)\ \textrm{case}\ \textrm{C28}$}
    \end{overpic}\\
    \vspace{2mm}
    \begin{overpic}[width=0.4\textwidth]
    {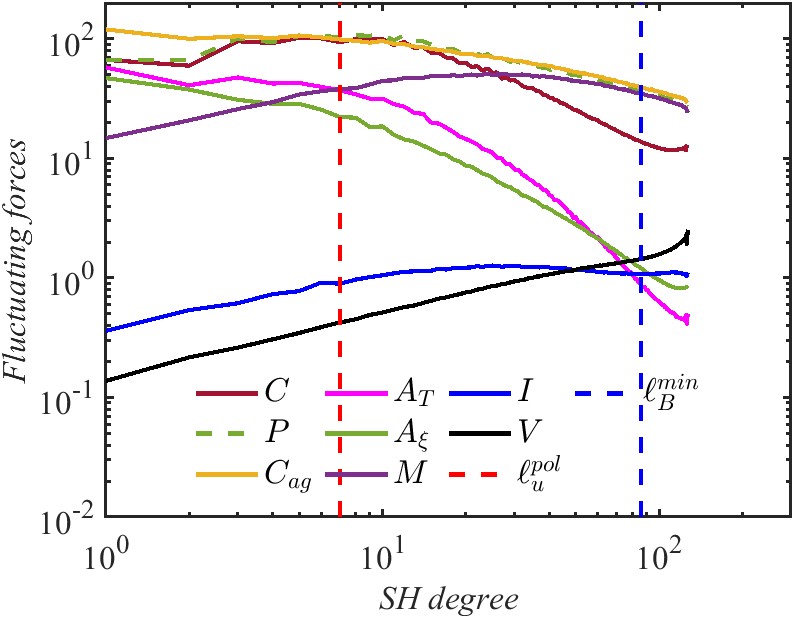}
    \put(0,80){$(b)\ \textrm{case}\ \textrm{DC05}$}
    \end{overpic}
    \caption{Force spectra for three example cases (a) C03 (no-magnetic without F-layer)(b) C28 (non-magnetic with F-layer) and (c) DC05 (dynamo with F-layer). Here $\lupol$ corresponds to the length scale associated with the peak of the poloidal kinetic energy and $\lbmin$ is the characteristic length scale of ohmic dissipation.}
    \label{SI:fspec}
\end{figure}

\section{Profile of radial magnetic field at the surface and core-mantle boundary}
\begin{figure}
    \centering
    \includegraphics[width=0.9\textwidth]{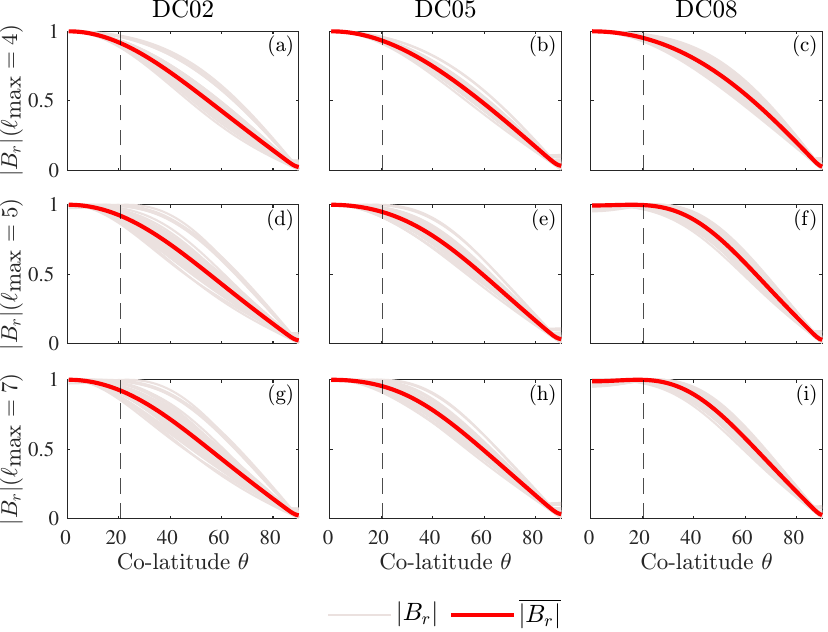}
    \caption{Latitudinal profiles of $|B_r|$ at the surface of the Earth, averaged in time and azimuth and between northern and southern hemispheres for three simulations: DC02 (a,d,g); DC05 (b, e, h); and DC08 (c, f, i), with a truncation degree $\ell_{\rm max}$ of 4, 5, and 7 (rows). Profiles are normalised by their maximum. Reddish grey lines show 100 snapshots of the azimuthally averaged profile of $|B_r|$. The F-layer cases have $r_s=0.83$. The latitude of the tangent cylinder is highlighted by the dashed vertical line. }
    \label{SI:mag_brprof_surf}
\end{figure}

\begin{figure}
    \centering
    \includegraphics[width=0.9\textwidth]{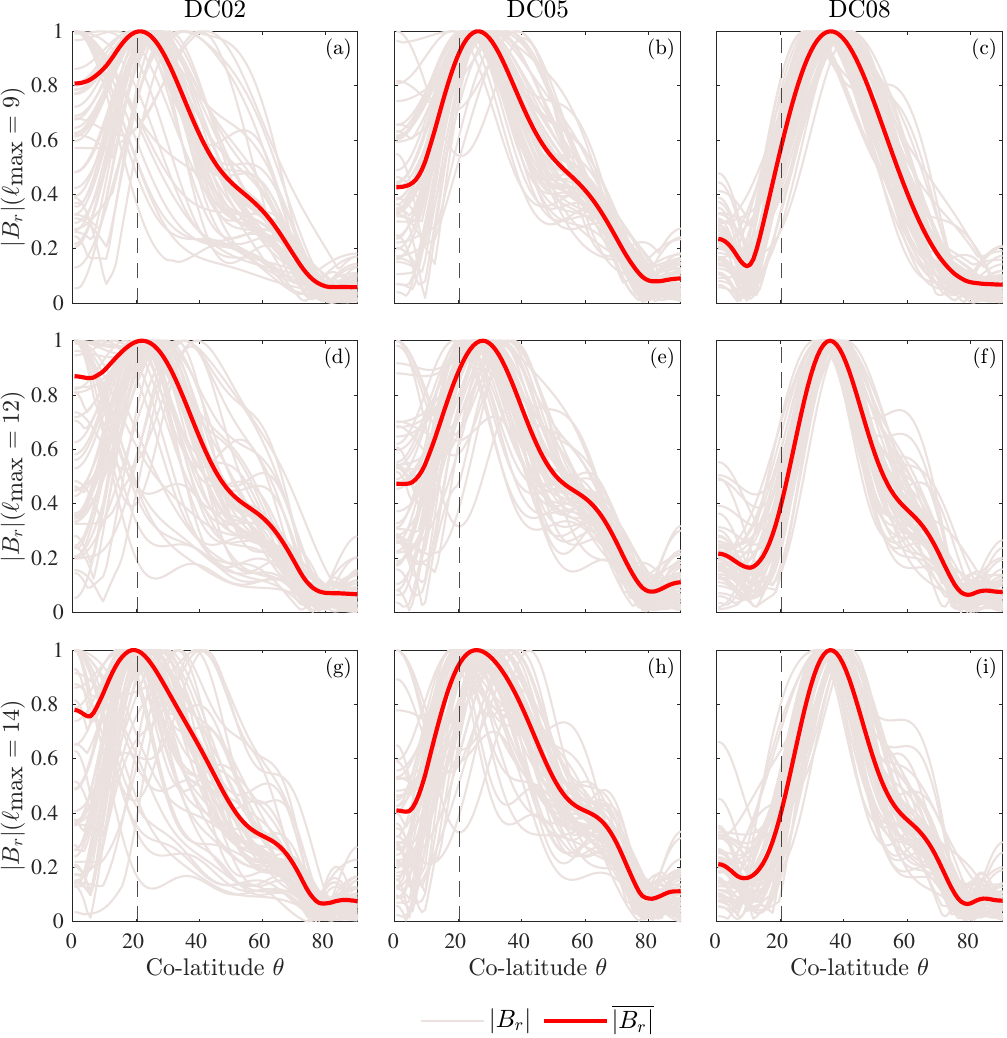}
    \caption{Latitudinal profiles of $|B_r|$ at $\ro$ averaged in time and azimuth and between northern and southern hemispheres for three simulations: DC02 (a,d,g); DC05 (b, e, h); and DC08 (c, f, i), with a truncation degree $\ell_{\rm max}$ of 9, 12, and 14 (rows). Profiles are normalised by their maximum. Reddish grey lines show 100 snapshots of the azimuthally averaged profile of $|B_r|$. The F-layer cases have $r_s=0.83$. The latitude of the tangent cylinder is highlighted by the dashed vertical line. }
    \label{SI:mag_brprof14}
\end{figure}

Figs. \ref{SI:mag_brprof_surf} and \ref{SI:mag_brprof14} are similar to Fig. \ref{fig:mag_brprof} in the main text, but at the surface of Earth for $\ell_{\textrm{max}}=4,\, 5,\, 7$ or at the core-mantle boundary with $\ell_{\textrm{max}}=9,\, 12,\, 14$.

\clearpage
\newpage
\bibliographystyle{cas-model2-names}
\bibliography{biblio}

\printcredits

\end{document}